\documentclass[a4article,11pt]{article}
\usepackage[utf8]{inputenc}

\usepackage[sorting=none]{biblatex}
\usepackage{amsmath}
\usepackage{amssymb}
\usepackage{physics}
\usepackage{bm}
\usepackage{hyperref}
\usepackage{graphicx}
\usepackage[dvipsnames]{xcolor}

\DeclareMathOperator{\sgn}{sgn}
\newcommand{\upd}[1]{^\mathrm{#1}}
\newcommand{\ind}[1]{_\mathrm{#1}}

\newcommand{\rapidity}{\theta}
\newcommand{\rapidityp}{\alpha}

\newcommand{\quasiparticleop}{\hat{\rho}}

\newtheorem{remark}{Remark}

\newcommand{\bc}{\begin{center}}
\newcommand{\ec}{\end{center}}
\def\ba#1{\begin{array}{#1}\displaystyle}
\newcommand{\ea}{\end{array}}

\newcommand{\beq}{\begin{equation}}
\newcommand{\eeq}{\end{equation}}
\newcommand{\beqa}{\begin{eqnarray}}
\newcommand{\eeqa}{\end{eqnarray}}
\newcommand{\no}{\nonumber}
\newcommand{\n}{\nonumber\\}
\newcommand{\bi}{\begin{itemize}}
\newcommand{\ei}{\end{itemize}}

\def\t#1{\tilde{#1}}
\def\h#1{\hat{#1}}
\def\b#1{\bar{#1}}
\def\frc#1#2{\frac{#1}{#2}}

\newcommand{\p}{\partial}

\newcommand{\N}{{\mathbb{N}}}
\newcommand{\R}{{\mathbb{R}}}

\newcommand{\ep}{\epsilon}

\newcommand{\ri}{{\rm i}}

\newcommand{\1}{{\bf 1}}

\DeclareMathOperator{\He}{He}
\def\bs#1{\boldsymbol{#1}}
\newcommand{\nor}{{\bs{:}}}

\newcommand{\change}[1]{#1}

\begin{document}


\begin{center}
{\Large\bf Towards an ab initio derivation of generalised hydrodynamics
from a gas of interacting wave packets}

\medskip
Benjamin Doyon, Friedrich H\"ubner

\medskip

Department of Mathematics, King's College London, London WC2R 2LS, U.K.

\end{center}






\begin{abstract}
We present steps towards an ab initio derivation of generalised hydrodynamics in quantum integrable models, starting from the Bethe wave functions, and explained on the example of the repulsive Lieb-Liniger model. This includes an identification of the generalised hydrodynamics quasi-particles as wave packets in the quantum model. These wave packets evolve according to a classical particle model and collect two-particle scattering shifts similar to solitons in integrable PDEs. We then discuss potential routes to obtain the generalised hydrodynamics equation for average conserved densities in long-wavelength states from this description. As part of this, we provide an explicit formula for the action of the spectral phase-space density operator on Bethe wave functions, and show that it generates local conserved densities.
\end{abstract}

\tableofcontents

\section{Introduction}

In recent years there has been much interest in developing hydrodynamic descriptions of many-body quantum systems in low dimensions \cite{jphysAspecialhydro}. One such avenue is generalised hydrodynamics (GHD) \cite{Doyon2016,Bertini2016}, a hydrodynamic theory for integrable systems, which accounts for the presence of an extensive number of conserved quantities. GHD applies widely, from ``integrable turbulence" \cite{zakharov2009turbulence}, including gases of solitons \cite{El_2021,suret2023pre}, and other classical systems \cite{Doyon_2017,PhysRevLett.120.164101,Bastianello2018,spohn_generalized_2019,doyon_generalised_2019,Bulchandani_2019,kochExact2023}, to quantum spin chains, field theories and gases \cite{Doyon2016,Bertini2016,PhysRevLett.119.195301,Bulchandani2017,Bulchandani2018,10.21468/SciPostPhys.6.6.070,unstable}. GHD has been confirmed experimentally in cold atoms \cite{Schemmer2019,Malvania2020,Moller2021}, and has seen far-reaching theoretical developments, see the reviews \cite{Doyon2020Notes,specialissueGHD,spohnbook,ESSLER2022127572,Bouchoule_2022,watsonGHDreview}.

At the Euler scale -- where viscous and diffusive effects are scaled away -- GHD takes a universal structure. The spectral phase-space density $\rho_{\rm p}(\theta,x,t)$ representing the local state of the system at time $t$ and position $x$, as a function of the ``rapidity", or spectral parameter, $\theta$, satisfies the conservation law \cite{Doyon2016,Bertini2016}
\beq\label{ghd}
	\p_t \rho_{\rm p}(t,x,\theta,x) + \p_x \big(v^{\rm eff}(t,x,\theta)\rho_{\rm p}(t,x,\theta)\big) = 0.
\eeq
The current in the second term is set by an ``effective velocity" $v^{\rm eff}(t,x,\theta) = v^{\rm eff}_{[\rho_{\rm p}(t,x,\cdot)]}(\theta)$, which depends functionally and nonlinearly on the state at $(x,t)$. In its simplest form, the effective velocity, for some state $\rho_{\rm p}(\theta)$, is the solution to the linear integral equation 
\beq\label{veff}
	v^{\rm eff}_{[\rho_{\rm p}]}(\theta) = v(\theta) + \int \dd \alpha\,
	\varphi(\theta-\alpha)\rho_{\rm p}(\alpha)(v^{\rm eff}_{[\rho_{\rm p}]}(\alpha) - v^{\rm eff}_{[\rho_{\rm p}]}(\theta))
\eeq
where $v(\rapidity)$ is the bare single particle velocity and $\varphi(\theta-\alpha)$  is the two-body semi-classical scattering shift, which encodes the interaction (see below).

Physically, $\rho\ind{p}(t,x,\theta)$ is a density in the phase space of asymptotic excitations (``quasi-particles") and their momenta $\theta$ (``rapidity"): $\dd \theta\dd x\,\rho\ind{p}(t,x,\theta)$ is the number of particles with asymptotic momenta in $[\theta,\theta+\dd\theta]$ seen in a time-of-flight thought experiment, where the ``fluid cell" at $x,t$, of mesoscopic length $\dd x$, is taken out of the fluid and let to expand on the line $x\in\R$, in the vacuum. Depending on the model to which GHD is applied, asymptotic excitations may be actual particles, classical solitons, stable quantum quasi-particle excitations, radiative waves, etc. $\rho\ind{p}(t,x,\theta)$ is well defined in non-integrable models as well, where the dependence on $\theta$ is restricted to particular functions (e.g.~the black-body radiation) \change{because of local relaxation at the basis of Euler hydrodynamics}. In integrable models, instead, the dependence on $\theta$ is essentially arbitrary, \change{even under local relaxation}, because all momenta are preserved in scattering events (elastic scattering) \cite{Smirnov1992,mussardo2010statistical}. We note that it is now possible \cite{Malvania2020}, by time-of-flight methods in cold atom gases, to experimentally access (an approximation of) $\rho\ind{p}(t,x,\theta)$.

Where does \eqref{ghd} come from? In the quantum context, beyond non-interacting models \cite{fagottiSciPost,granetGHDfree}, GHD is derived phenomenologically \cite{Doyon2016,Bertini2016}. Like for most hydrodynamic equations, this relies on the {\em hydrodynamic assumption}: in every fluid cell, entropy is independently maximised with respect to the available extensive conserved quantities. Euler-scale hydrodynamic equations are the associated conservation laws, written under this assumption. With integrability, there are infinitely many, and generalised Gibbs ensembles  (GGE; see e.g.~\cite{Calabrese_2016}) are the entropy-maximised states. The Bethe ansatz structure allows one to recast the infinitely-many equations into \eqref{ghd}; expressions for average currents (first derived in \cite{Doyon2016}) are the crucial results, see the reviews \cite{borsi2021current,cubero2021form}. 

These derivations are by now standard and have been carried out for many quantum integrable models, see e.g.~the reviews in the special issue \cite{specialissueGHD}. However, they are based on special results from integrability such as the thermodynamic Bethe ansatz (TBA), and, besides the local equilibrium paradigm of hydrodynamics, they do not  provide much intuition behind the origin of \eqref{ghd} and with \eqref{veff}. By contrast, Eqs.~\eqref{ghd} and \eqref{veff}  themselves have a very intuitive, kinetic interpretation, inspired by factorised scattering theory \cite{ZAMOLODCHIKOV1979253,novikov1984theory,Faddeev:1987ph,Smirnov1992,mussardo2010statistical}, proposed in \cite{PhysRevLett.95.204101} for soliton gases, and in \cite{Doyon2018sol} for quantum integrable models. Indeed, the evolution of the spectral phase-space density is governed by processes whereby quasi-particles, seen as actual particles within the gas, travel freely at velocity $v(\theta)$ except for jumps --  spatial Wigner shifts -- of size $\varphi(\theta-\alpha)$ at collisions with other particles. The accumulation of a macroscopic amount of jumps modifies their overall velocities, and this is what sets the spectral phase-space current. The derivations of Eqs.~\eqref{ghd} and \eqref{veff} currently available in quantum models do not explain how this simple picture emerges, even though the effective velocity $v^{\rm eff}$ itself arises naturally from the Bethe ansatz as an interacting ``group velocity'' \cite{PhysRevLett.113.187203}. Further, they have some logical flaws: for instance, one typically assumes periodic boundary conditions of fluid cells to compute expectation values inside them using the TBA. While for sufficiently large fluid cells boundary conditions should not matter, this is clearly not physical: neighboring fluid cells are connected to each other, not to themselves. Also, these periodic boundary conditions introduce a spurious quantitization of momenta, which eventually gives rise to the interaction term in \eqref{ghd}. In truth, the system can be on the infinite line, where any momentum quantization is simply absent. Finally, even if these logical flaws are ignored, it is not clear how well the local relaxation assumption of hydrodynamics works: since fundamentally a pure initial state has to remain pure, local relaxation can only occur for local observables. While all of this is believed to be not important on the Euler scale, the precise nature of the emergence of hydrodynamics might affect higher order corrections like diffusion. 

\change{The goal of this paper is to provide steps towards alternative derivations of GHD starting from the microscopic dynamics, avoiding the assumption of local equilibrium. We will present {\em four separate proposals} that can each, if completed, give rise to a full derivation of the GHD equations. As it currently stands, in each proposal some steps are missing, which we will point out. It is startling that ten years on from the inception of GHD there are, as far as we are aware, no other derivation, even at the physical level of rigour, in interacting quantum systems.}

\change{Derivations of GHD equations from the microscopic dynamics} exist in many classical models, for instance hard-rod gases \cite{Boldrighini1983,ferrariHR2022} and certain cellular automata \cite{ferrari_nguyen_rolla_wang_2021,2020cs} by using ``contraction maps'', and in soliton gases by using Whitham modulation theory \cite{el_thermodynamic_2003,el2020spectral}. The main idea put forward here can be seen as a version of the contraction map, and a version of modulation theory. 

\change{In the first two proposals we consider an explicit large-scale analysis of the time-dependent wave function expressed using Bethe eigenfunctions, and the consequences these have on averages of local conserved densities. The first is based on a stationary phase analysis similar to the semi-classical limit of quantum mechanics, Sec.~\ref{ssectapprox1} and \ref{ssectghdwavefunction}, and the second, on a large-scale modulation of Bethe wavefunction, Sec.~\ref{ssectapprox2} and \ref{ssectghdlocalequil}. In the third, we directly consider averages of local conserved densities via an interacting generalisation of the Wigner function, Sec.~\ref{secwig}. Finally, in the fourth, we consider the Heisenberg time evolution of position and rapidity operators defined via the Bethe eigenfunctions, and its consequences on averages of local conserved densities, Sec.~\ref{sec:heisenberg}.
}

\change{In all derivations, we obtain {\em an intermediate model of classical particles}.} The classical particle model, that we refer to as ``semi-classical Bethe model'', is a generalisation of the contraction maps used in classical settings. GHD then emerges as the large-scale theory for this model. We carry out this derivation on the example of repulsive Lieb-Lingier model, but it can easily be adapted to other quantum integrable models. One aspect of these results is that they suggest not only the kinetic origin of the GHD equation in a quantum model, but also a microscopic understanding of the wave function itself and its evolution in time. The stationary phase and modulation analyses also give a clear explanation for why it is the semi-classical scattering shift $\varphi(\rapidity) = \phi'(\rapidity)$, the derivative of the quantum scattering phase, that enters the effective velocity equation \eqref{veff}. \change{Note that similar techniques to extract  effective particle dynamics in quantum systems are well established in non-interacting models~\cite{PhysRev.76.1592,PhysRev.84.814,PhysRevB.59.14915,fagottiSciPost}. In this sense our work provides steps towards extensions of these methods to the interacting case.}

Our work starts with the Bethe eigenfunctions of the Lieb-Liniger model \cite{PhysRev.130.1605}, Eq.~\eqref{equ:LL_Hamiltonian} below:
\begin{align}\label{equ:wavefctintro}
    \psi(\bs{x}|\bs{\rapidity}) &= c_Ns(\bs{x})\sum_{\sigma\in \mathcal{S}_N} (-1)^{|\sigma|} e^{\ri\Phi(\bs{x}_{\sigma},\bs{\rapidity})}.
\end{align}
$s(\bs{x})$ and $c_N$ are given below Eq.~\eqref{equ:pre_int_LL_Nparticle}, and the Bethe phase $\Phi(\bs{x},\bs{\rapidity})$ by \eqref{equ:pre_int_LL_phase}. 

\change{The main derivations are organised into three (connected) sections: the first is an analysis of wave functions within the Schroedinger picture of the Lieb-Liniger model, the second an analysis of expectation values of conserved densities, and the latter an analysis of operators within the Heisenberg picture:
\begin{itemize}
    \item In Sec.~\ref{sec:derivation_wavepackets} we explain our first two proposals by studying the evolution of the wave function, and argue that it can be approximated as a superposition of classically evolving wave packets: the effective position of a wave packet $x_i(t)$, in time, evolves according to
\begin{align}
    y_i^0 + \rapidity_i t &= x_i(t) + \frac{1}{2} \sum_{i\neq j} \sgn(x_i(t)-x_j(t))\varphi(\rapidity_i-\rapidity_j)
    \label{equ:LL_SP_QP_def_micro_intro}
\end{align}
where $y_i^0$ encode their initial positions (simply by setting $t=0$ in this equation). In particular, in the large scale limit the velocities of these particles are given by the effective velocity Eq.~\eqref{veff}. This clarifies for the first time the underlying kinetic interpretation of \eqref{ghd} in a quantum model.

Furthermore, we show that \eqref{equ:LL_SP_QP_def_micro_intro} has a unique solution for $\varphi(\rapidity)$ chosen to be the two-body semi-classical scattering shift of the Lieb-Liniger model, obtained by the minimisation of a convex function.

\item In Sec.~\ref{sec:expecation_values} we study the evolution of expectation values (which is the standard problem considered in GHD). We add to the first two proposals by applying the results for wave functions to expectation values. We then introduce our third proposal, that of the interacting Wigner function.

The derivations in this section are only possible due to an important new result: we obtain an explicit formula for {\em the action of the local densities $\hat{q}_a(x)$ of conserved quantities $\hat{Q}_a= \int\dd{x}\hat{q}_a(x)$ on Bethe eigenstates \eqref{equ:wavefctintro}},  as follows:
\begin{align}\label{equ:qactionintro}           \hat{q}_a(x)\psi(\bs{x}|\bs{\rapidity}) &= c_Ns(\bs{x})\sum_{\sigma\in \mathcal{S}_N} (-1)^{|\sigma|} \qty[\sum_i \delta(x-x_{\sigma_i})\rapidity_i^a] e^{i\Phi(\bs{x}_{\sigma},\bs{\rapidity})} & a&\in\N.
\end{align}
This is a natural expression, but most importantly, {\em we show that these are local observables, supported at $x$}. Further, we define the {\em spectral phase-space density operator}
\beq\label{densityoperatorintro}
    \h\rho(x,\theta)=\sum_i \delta(x-\h x_i)\delta(\theta-\h \theta_i),
\eeq
whose action on Bethe eigenstates is
\begin{align}
    \quasiparticleop(x,\rapidity)\psi(\bs{x}|\bs{\rapidity}) &= c_Ns(\bs{x})\sum_{\sigma\in \mathcal{S}_N} (-1)^{|\sigma|} \qty[\sum_i \delta(x-x_{\sigma_i})\delta(\rapidity-\rapidity_i)] e^{i\Phi(\bs{x}_{\sigma},\bs{\rapidity})}\label{equ:LL_deriv_cq_def_intro}
\end{align}
and which generates the local densities as
\beq
	\hat q_a(x) = \int \dd\theta \,\theta^a \h\rho(x,\theta).
\eeq
The GHD equation is an equation for its average $\rho\ind{p}(x,\theta) = \langle \h\rho(x,\theta)\rangle$ in long-wavelength thermodynamic states $\langle \cdots\rangle$.

\item In Sec. \ref{sec:heisenberg} we introduce our fourth proposal: we investigate the emergence of GHD from a Heisenberg picture viewpoint.

The Heisenberg picture leads to a quantized version of the emergent semi-classical dynamics. With the usual quantum mechanical position operators $\hat x_i$, and natural rapidity operators $\hat \theta_i$ defined as being diagonal on the basis of Bethe ansatz scattering waves, we show that, under Heisenberg evolution, Eq.~\eqref{evoly} holds:
\beq
    \hat y_i + \hat \theta_i t
    =\hat x_i(t) +
    \frc12 \sum_{j\neq i}
    \sgn(\hat x_i(t)-\hat x_j(t))\varphi(\hat \theta_i-\hat \theta_j)
\eeq
where the operators  $\h y_i$  encode the operatorial initial conditions, and are defined by setting $t=0$ in this equation. We pass from this operatorial equation to an equation for the empirical density operator \eqref{equ:LL_deriv_cq_def_intro} by using the height-field methods. These results are rigorous, and involve a particular normal ordering. We explain how, if the effect of normal ordering can be neglected on large scales of space, then this would naturally lead to the GHD equation: we argue for this by showing that fluid-cell means of charge densities commute in the macroscopic limit, suggesting the emergent classicality of GHD.
\end{itemize}
}

\medskip
{\em \noindent {\bf Note added (2nd version, August 2025)}:

In the first version of this manuscript (July 2023), we claimed to have found an ab initio derivation of GHD in the Lieb-Liniger model. Over time we got increasingly uncertain of the validity of the (many) assumptions used in there, and we concluded that we could not yet claim to have full derivation, even at the physical level of rigour. Some of the assumptions made would require a much better understanding.

Instead, we have thus decided to both highlight our intuitive understanding and to discuss various approaches we tried to turn intuition into formulas. The fundamental bottleneck seems to be that we try to apply well established mathematical methods tailored towards classical models on quantum models. In particular, the sum over permutations, such as that in \eqref{equ:qactionintro}, would need a better treatment when the large-scale asymptotic is taken.
We believe that once appropriate mathematical frameworks for quantum systems are developed, combining them with the ideas presented here will give much deeper insights into quantum integrable models, and eventually a proper ab initio derivation.

The first version of this manuscript is the first work where the particle system \eqref{equ:LL_SP_QP_def_micro_intro} appears in the literature. Since then, a number of works have developed aspects of it \cite{PhysRevLett.132.251602,doyon2023generalisedtbartdeformationsclassicalfree,Bonnemain2025,urilyon2025simulatinggeneralisedfluidsinteracting,aggarwal2025asymptoticscatteringrelationtoda,aggarwal2025effectivevelocitiestodalattice}, including our own. This system and its extensions are now seen as fundamental equations for the large-scale behaviour of integrable models, as we first proposed in \cite{PhysRevLett.132.251602,doyon2023generalisedtbartdeformationsclassicalfree} and developed in \cite{Bonnemain2025}. Notably, a recent work by A.~Aggarwal \cite{aggarwal2025asymptoticscatteringrelationtoda,aggarwal2025effectivevelocitiestodalattice} showed that \eqref{equ:LL_SP_QP_def_micro_intro} emerges as an equation for effective soliton centers in finite-density homogeneous soliton gases of the Toda model, as we predicted in \cite{PhysRevLett.132.251602}.}

\section{The Lieb-Liniger model}
The Lieb-Liniger model describes a gas of $N$ bosonic particles on the line with contact interaction~\cite{PhysRev.130.1605},
\beq
    \h H = -\frac{1}{2}\sum_{j=1}^N \frc{\p^2}{\p x_j^2} + \frac{c}{2} \sum_{(i \neq j) =1}^N
     \delta(x_i-x_j).\label{equ:LL_Hamiltonian}
\eeq
This Hamiltonian acts on the Hilbert space of fully symmetric square integable functions $L^2\ind{sym}(\R^N)$.

In this paper we study the much simpler repulsive case $c>0$, due to the absence of string states; however the techniques are general, and in appendix \ref{app:attractive} we briefly discuss the attractive case $c<0$. 

Using the notation $\bs x = (x_i)_{i\in \{1,\ldots,N\}}$, it is well known that the eigenstates of \eqref{equ:LL_Hamiltonian} are given by the coordinate Bethe ansatz~\cite{PhysRev.130.1605,jscaux}
\begin{align}
	\psi(\bs{x}|\bs{\rapidity}) = c_N s(\bs{x}) \sum_{\sigma\in \mathcal{S}_N} (-1)^{|\sigma|} e^{\ri\Phi(\bs{x}_{\sigma},\bs{\rapidity})},\label{equ:pre_int_LL_Nparticle}
\end{align}
where $s(\bs{x})=\prod_{i<j} \sgn(x_i-x_j)$, $c_N=1/\sqrt{(2\pi)^NN!}$, $\sigma$ runs over all permutations of $N$ elements, $\bs x_{\sigma}$ is the vector $(x_{\sigma_i})_i$, and the Bethe phase is given by
\begin{align}
	\Phi(\bs{x},\bs{\rapidity}) = \sum_i \rapidity_i x_i + \frac{1}{4} \sum_{i\neq j} \sgn(x_i-x_j) \phi(\rapidity_i-\rapidity_j).\label{equ:pre_int_LL_phase}
\end{align}
Here, $\phi(\rapidity) =2\arctan(\rapidity/c)$ is the two-particle scattering phase. The energy of this eigenstate is given by $E=\tfrac{1}{2}\sum_i\rapidity_i^2$. Note $\psi(\bs{x}|\bs{\rapidity})$ is explicitly symmetric in $\bs x$ (bosonic statistics), and antisymmetric in $\bs\rapidity$. In particular, the wave function vanishes whenever any $\rapidity_i = \rapidity_j\,(i\neq j)$. This is interpreted as a Pauli exclusion principle: even though the physical particle is bosonic, the quasi-particle has emergent fermionic statistics because of the contact interaction. For $\bs\theta\in \R^N_> = \{\bs\theta:\theta_1>\theta_2>\ldots>\theta_N\}$, these eigenstates form an orthonormal scattering basis of $L^2\ind{sym}(\mathbb{R}^N)$~\cite{gaudinbook}
\begin{align}
    \int\dd[N]{x}\psi^*(\bs{x}|\bs{\rapidity})\psi(\bs{x}|\bs{\rapidityp}) &= \prod_i\delta(\rapidity_i-\rapidityp_i),\quad \bs\rapidity,\bs\rapidityp\in\R^N_>,\label{equ:pre_int_LL_basis_1}\\
    \int_{\R^N_>}\dd[N]{\rapidity}\psi(\bs{x}|\bs{\rapidity})\psi^*(\bs{x}'|\bs{\rapidity}) &= \frc1{N!}\sum_{\sigma \in \mathcal{S}_N} \prod_i\delta(x_i-x'_{\sigma_i}).\label{equ:pre_int_LL_basis_2}
\end{align} 

The GHD equation of this model has been developed in~\cite{Doyon2016} and is given by \eqref{ghd} with $v(\theta) = \theta$ and $\varphi(\rapidity) = \phi'(\rapidity)$. These results were derived using the TBA formalism, which allows for the computation of thermodynamic quantities in integrable systems. For technical reasons\footnote{Thermodynamics studies homogeneous finite density states, which do not exist on the infinite line (no matter how many particles are added, the system always has zero density).} one has to consider the system in a finite (say periodic) box of size $\ell$ and then send $N\sim \ell \to \infty$. While in the infinite system all states like \eqref{equ:pre_int_LL_Nparticle} are eigenstates, on a finite system only $\bs{\theta}$ which satisfy the so-called Bethe quantization conditions are eigenstates. These conditions are highly non-linear equations and no general explicit solutions are known (certain solutions are obtained in special limits, see e.g.~\cite{miao2021exact}). In the TBA formalism, as $N\sim \ell \to \infty$, one then carries out a continuum limit of the Bethe quantization conditions. In this limit one finds a non-trivial density of states. This procedure is the mathematical origin of the fact that the GHD scattering shift is given by the derivative $\varphi(\rapidity) = \phi'(\rapidity)$ of the two particle scattering phase. However, this does not provide the kinetic interpretation of the GHD equation discussed in the introduction.

One of the advantages of the derivation presented here is that it circumvents the need to put the system in a finite box. We start directly from the infinite system. Thus, no Bethe quantization is required, and we will see that the identity $\varphi(\rapidity) = \phi'(\rapidity)$ naturally emerges from a stationary-phase, large-scale analysis.

\section{Gas of wavepackets and slowly-varying amplitude modulations in the Lieb-Liniger model}\label{sec:derivation_wavepackets}

The starting point for our derivation is the decomposition of any wavefunction into eigenstates of the Lieb-Liniger model
\begin{align}
	\Psi(t,\bs{x}) &= \int\dd[N]{\rapidity} A(\bs{\rapidity}) \psi(\bs{x}|\bs{\rapidity}) e^{-\tfrac{\ri}{2}\sum_i \rapidity_i^2 t}.\label{equ:LL_SP_psi_starting_point}
\end{align}
Here, the amplitude $A(\bs{\rapidity})$ can be an arbitrary (suitably normalized) square integrable function. Only its antisymmetric part will contribute due to the antisymmetry of $\psi(\bs{x}|\bs{\rapidity})$ in $\bs\rapidity$.

\change{We consider two setups. In the first setup, we make the following intuitive choice, which we will call {\em gas of wave packets}
\begin{align}
	A(\bs{\rapidity}) &= \prod_{i=1}^N {A}_i\qty(\tfrac{\rapidity_i-\rapidity^0_i}{\Delta \rapidity}) \,e^{-\ri\sum_i \ell \rapidity_i y_i^0}.\label{equ:LL_SP_psi_amplitude}
\end{align}
Each particle is described by an independent wave packet of shape ${A}_i(\rapidity)$, shifted to the rapidity $\rapidity_i^0$ and rescaled to a width $\Delta \rapidity=\Delta\theta(\ell)$. Here, the parameter $\ell \gg 1$ is the large, macroscopic length scale, which controls both {\em the hydrodynamic limit} and {\em the wave packet sizes and positions}, and again the functions ${A}_i(\rapidity)$ do not depend on $\ell$. For instance, one could choose $A_i(\rapidity)$ to be identical Gaussians
\begin{align}
    A_i(\rapidity) = e^{-\rapidity^2/2},
\end{align}
but much more general choices are possible. As above, the initial position is controlled by the oscillatory behaviour of the wave packets; ${A}_i(\rapidity)$ may be oscillatory, but in the limit where $\ell$ is large, this is a microscopic effect, and the macroscopic positions of the wave packets will be controlled by the fast oscillation $\ell y_i^0$. The values of $y_i^0$ will not turn out to be the actual positions of the wave packets, but rather their locations in the contracted space (see discussion in Sec.~\ref{sec:soliton_gas_interpretation}). The width of the wave packets in position space is $\Delta x_i \propto 1/\Delta \rapidity$. We choose $\Delta\rapidity = \Delta\rapidity(\ell)$ as
\beq\label{equ:width}
    1/\ell \leq \Delta \rapidity \leq 1.
\eeq
Since we want to describe a finite density state, we will study states where
\beq\label{equ:macrolimit}
    \ell,N\to\infty,\quad N \propto \ell
\eeq
and where the distribution of $(y_i^0,\rapidity_i^0)$ approaches a continuous, normalisable distribution ${\t \rho}^0(y^0,\rapidity^0)$. 

In real space, the wave function \eqref{equ:LL_SP_psi_starting_point} for the gas of wave packets, evaluated at macroscopic coordinate $x$, is
\beq\label{Psimodulatedinteracting}
    \Psi(0,\ell \bs x) = \t c_Ns(\bs x)\sum_{\sigma\in S_N}
    (-1)^{|\sigma|}
    \t A(\bs x_\sigma,\bs y^0)
    e^{\ri\ell\b \Phi(\bs x_\sigma,\bs\theta^0)
    }
\eeq
with the normalisation $\t c_N = e^{-\ri \ell\bs\theta^0\cdot \bs y^0} (\Delta\theta)^N c_N$, and where
\beq\label{modulatingfctinteractingA}
    \t A(\bs x,\bs y^0) =
    \int \dd[N]{\alpha}
    \prod_i A_i(\alpha_i)
    e^{\ri \ell (
    \b\Phi(\bs x,\bs\theta^0 + \Delta\theta\bs\alpha)
    -
    \b\Phi(\bs x,\bs\theta^0)
    -\Delta\theta\bs\alpha \cdot \bs y^0)}.
\eeq
Here we introduced the scaled Bethe phase
\beq
	\b\Phi(\bs{x},\bs{\rapidity}) = 
    \frc1\ell \Phi(\ell\bs{x},\bs{\rapidity}) = 
    \sum_i \rapidity_i x_i + \frac{1}{4\ell} \sum_{i\neq j} \sgn(x_i-x_j) \phi(\rapidity_i-\rapidity_j).\label{Phibar}
\eeq
Note that in the limit \eqref{equ:macrolimit}, the scaled Bethe phase becomes a sum of well-defined phases for each coordinates, in term of the normalised distribution $\rho(x,\theta) = \ell^{-1}\sum_i \delta(x-x_i)\delta(\theta-\theta_i)$:
\beq\label{phibarlimit}
    \b\Phi(\bs{x},\bs{\rapidity}) \to 
    \sum_i \Big(\theta_i x_i
    +\frc14
    \int \dd{x}\dd{\theta}\rho(x,\theta)
    \sgn(x_i-x)\phi(\theta_i-\theta)
    \Big).
\eeq

In the second setup, we consider instead a choice that is more physically transparent in the real-space form of the wave function. In this setup, the initial wave function has the following form in terms of the macroscopic spatial coordinate $x$:
\beq\label{Psimodulatedinteractinggeneral}
    \Psi(0,\ell \bs x) = c_Ns(\bs x)\sum_{\sigma\in S_N}
    (-1)^{|\sigma|}
    B(\bs x_\sigma)
    e^{\ri\ell\b \Phi(\bs x_\sigma,\bs\theta^0)
    }.
\eeq
The ``modulating function'' $B(\bs x)$ is required to approach, in some sense, a regular enough function of each of its argument $x_i$ in the limit \eqref{equ:macrolimit}, up to a phase and normalisation. Eq.~\eqref{Psimodulatedinteractinggeneral} is therefore a {\em slowly-varying amplitude modulation}, where the interacting Bethe wave $e^{\ri\ell\b \Phi(\bs x,\bs\theta^0)}$ oscillates on a microscopic scale, and is modulated by $B(\bs x)$, which varies on the much larger macroscopic scales. Then, this form of the initial wave function looks, locally, like an eigenfunction. In terms of the general form \eqref{equ:LL_SP_psi_starting_point}, this means that we choose $A(\bs\theta)$ dependent on the macroscopic parameter $\ell$ in such a way that, in the limit \eqref{equ:macrolimit}, the function
\beq\label{defB}
    B(\bs x) =(\Delta\theta)^N
    \int \dd[N]{\alpha}
    A(\bs\theta^0 + \Delta \theta\bs\alpha)
    e^{\ri \ell (
    \b\Phi(\bs x,\bs\theta^0 + \Delta\theta\bs\alpha)
    -
    \b\Phi(\bs x,\bs\theta^0)
    )},
\eeq
with
\beq
    \Delta \theta =1/\ell,
\eeq
approaches a regular enough function.

We do not attempt to give a precise meaning to ``regular enough'', but provide two examples. An example of modulating function $B(\bs x)$ is the factorised form
\beq\label{Bfacto}
    B(\bs x) \sim \prod_i B_i(x-x_i^0)
\eeq
with $B_i(\bs z)$ independent of $\ell$, and with $(x_i^0,\theta_i^0)$ approaching a continuous, normalisable distribution $\rho^0(x^0,\theta^0)$. Another example is
the gas of wave packet of the first setup, with the particular choice $\Delta\theta = 1/\ell$: for that example, \eqref{Bfacto} does not hold, but instead
\beqa
    B(\bs x) &=&
    e^{-\ri \ell\bs\theta^0\cdot \bs y^0} \ell^{-N}
    \int \dd[N]{\alpha}
    \prod_i A_i(\alpha_i)
    e^{\ri \ell (
    \b\Phi(\bs x,\bs\theta^0 + \ell^{-1}\bs\alpha)
    -
    \b\Phi(\bs x,\bs\theta^0)
    -\ell^{-1}\bs\alpha \cdot \bs y^0)}
    \n
    &=&
    e^{-\ri \ell\bs\theta^0\cdot \bs y^0} \ell^{-N}
    \t A(\bs x,\bs y_0)\big|_{\Delta\theta=1/\ell}.
    \label{modulatingfctinteracting}
\eeqa
Note that the overall phase $e^{-\ri \ell\bs\theta^0\cdot \bs y^0}$ here is not physically relevant, but only a consequence of our definitions. We see, by Taylor series expansion in the exponential and using \eqref{phibarlimit}, that the exponent in the exponential takes an $\ell$-independent form in the limit \eqref{equ:macrolimit}, hence the modulating function \eqref{modulatingfctinteracting} indeed has a regular enough form in the macroscopic limit \eqref{equ:macrolimit}.

It is instructive to illustrate the above setups in the case of a single, non-interacting particle
\beq
    \psi(x|\theta) = \frc1{\sqrt{2\pi}}e^{\ri \theta x}
\eeq
of momentum $\theta$ and energy $\theta^2/2$. This is \eqref{equ:pre_int_LL_Nparticle} with $N=1$. Then, a wave packet, controlled by the large parameter $\ell\gg 1$, is the decomposition \eqref{equ:LL_SP_psi_starting_point}
with
\begin{align}
	A(\rapidity) &= A_1\qty(\tfrac{\rapidity-\rapidity^0}{\Delta \rapidity(\ell)}) \,e^{-\ri \ell \rapidity y^0}
    \label{equ:LL_SP_psi_amplitude_free}
\end{align}
where $A_1(\theta)$ is an $\ell$-independent, well-behaved function, such as a Schwartz function, $\theta^0$ determines the position of the wave packet in momentum space, $\Delta \theta(\ell)$ its $\ell$-dependent spread, and $y^0$ a fixed parameter. In order to see what such a wave packet means in real space, consider a slowly-varying amplitude modulation of the eigenfunction: a wave function that takes the form, in terms of the scaled coordinates $x$,
\beq\label{equ:amplitudemodulationfree}
    \Psi(0,\ell x) = \frc1{\sqrt{2\pi}} B(x) e^{\ri \ell \theta^0 x}
\eeq
where $B(x)$, the modulating function, is again a well-behaved function, independent of $\ell$. Locally, on any small intervals of the scaled coordinates $x\in [a,a + b/\ell]$, this is a free wave function of momentum $\theta^0$, but the amplitude is set by $B(x)$, and hence depends on where the small interval lies. Writing its Fourier transform as
\beq
    B(x) = \frc{e^{-\ri \ell\theta^0y^0}}{\ell}\t A(x-y^0),\quad  \t  A(x) = \int \dd\alpha\,A_1(\alpha)e^{\ri \alpha x},
    \label{equ:Atilde}
\eeq
Eq.~\eqref{equ:amplitudemodulationfree} has the form \eqref{equ:LL_SP_psi_starting_point} (at $N=1$ and $t=0$) with \eqref{equ:LL_SP_psi_amplitude_free} and $\Delta\theta(\ell) = 1/\ell$. Therefore, a sharp wave packet around $\theta$ spanning momenta $\theta^0+\mathcal O(\ell^{-1})$ is an amplitude modulation \eqref{equ:amplitudemodulationfree}, where the fast-oscillatory part of the wave packet determines the positional shift of the modulating function. A scaling of $\Delta\theta(\ell)$ that decays slower than $1/\ell$ would give an amplitude modulation that is sharper in real space, of extent smaller that $\mathcal O(1)$ in scaled coordinate.

The form \eqref{modulatingfctinteracting} for the modulating function coming from the gas of wave packets is an immediate generalisation of \eqref{equ:Atilde}, with the difference that $\bs y_0$ is no longer simply a linear shift of $\bs x$. Then, much like the single-particle example above, wave packets are of macroscopic extent in real space.
For the gas of wave packet formulation, we may take larger values $\Delta\theta\gg 1/\ell$. In this case, wave packets
from \eqref{Psimodulatedinteracting}
give rise to point-like particles in real-space macroscopic coordinates, in the macroscopic limit \eqref{equ:macrolimit}. They are still large on microscopic scales, hence the wave function still looks locally like an eigenfunction, but the time evolution is quicker.

We will use two approximation schemes  in order to evaluate the dynamics in the two setups above. These two schemes form the basis of our first two proposals of derivations of the GHD equation. In the first, we use a stationary phase approximation, while in the second, we use a linearisation of the Bethe phase.

We will show, by linearisation of the Bethe phase, Subsec.~\ref{ssectapprox2}, that the slowly-varying amplitude modulation \eqref{Psimodulatedinteractinggeneral}, and the gas of wave packets \eqref{Psimodulatedinteracting} with $1/\ell\leq \Delta\theta\ll 1/\sqrt{\ell}$, evolve slowly, with coordinates of the modulating function, or position of the wave packets, satisfying {\em an effective interacting-particle equation}. This will give rise to the hydrodynamic behaviour that we expect at large scale. The amplitude modulation picture is similar to Whitham modulation theory used in integrable PDE's to derive GHD. We will also show, by stationary phase approximation Subsec.~\ref{ssectapprox1}, that time evolution of the gas of wave packets for the sharp case $1/\sqrt{\ell}\ll \Delta\theta\leq 1$ gives rise to {\em the same effective interacting-particle equation}, but now for the positions of the point-like wave packets seen on macroscopic scales.}

We discuss in more details in Sec.~\ref{sec:gas_discussion} the constraints on $\Delta\theta = \Delta \theta(\ell)$ for these two approximation schemes.

\subsection{Approximation Scheme 1: Stationary phase approximation}\label{ssectapprox1}

\change{The first approximation scheme is valid under the following assumption (see discussion in Sec.~\ref{sec:gas_discussion}):}
\beq\label{equ:widthmore}
    1/\sqrt{\ell} \ll \Delta \rapidity \leq 1.
\eeq
As mentioned, this means that in real space, under \eqref{equ:widthmore}, wave packets are supported on a length scale much smaller than $\sqrt{\ell}$, hence are point-like particles in macroscopic coordinates. We will now obtain the equation for these emergent classical point particles.

Going to macroscopic coordinates $x\to \ell x$ and $t \to \ell t$, we have from \eqref{equ:LL_SP_psi_starting_point} and \eqref{equ:LL_SP_psi_amplitude}
\begin{align}
	\Psi(\ell t,\ell\bs{x}) &= c_Ns(\bs x) \sum_{\sigma\in \mathcal{S}_N} (-1)^{|\sigma|} \int\dd[N]{\rapidity} e^{\ri\ell S_t(\bs{x}_{\sigma},\bs{y}^0,\bs{\rapidity})} \prod_i{A}_i\qty(\tfrac{\rapidity_i-\rapidity^0_i}{\Delta \rapidity}), \label{equ:LL_SP_psi_macroscopic}
\end{align}
where the phase is given by $S_t(\bs{x},\bs{y}^0,\bs{\rapidity}) = \b\Phi(\bs x,\bs\theta) - \bs\theta\cdot\bs y^0 - \frc12\bs\theta^2 t$, that is
\begin{align}
	S_t(\bs{x},\bs{y}^0,\bs{\rapidity}) &= \sum_i \rapidity_i (x_i-y_i^0) +\frac{1}{4\ell}\sum_{i\neq j}\sgn(x_i-x_j) \phi(\rapidity_i-\rapidity_j)- \frac{1}{2}\sum_i\rapidity_i^2t.\label{equ:LL_SP_phase}
\end{align}

We now consider the limit \eqref{equ:macrolimit}. As there are $N$ rapidity integrals, we consider the phase factor for each rapidity integral,
\[
    \rapidity_i (x_i-y_i^0) +\frac{1}{4\ell}\sum_{j\neq i}\sgn(x_i-x_j) \phi(\rapidity_i-\rapidity_j)- \frac{1}{2}\rapidity_i^2t.
\]
Note that the term on the right-hand side containing the sum is of order $\mathcal O(1)$, because there are $\sim N$ finite terms in the sum, with a factor $1/\ell$. Neglecting the potential effects of the large number of permutations over which we sum (see the discussion below), as $\ell \to \infty$ we approximate the  integrals, for fixed permutation $\sigma$, using the stationary phase approximation. That is, the dominant contribution should come from the rapidities satisfying
\begin{align}
    0 &= \partial_{\rapidity_i} S_t(\bs{x},\bs{y}^0,\bs{\rapidity}) = x_i-y_i^0 + \frac{1}{2\ell} \sum_{j\neq i}\sgn(x_i-x_j) \varphi(\rapidity_i-\rapidity_j)- \rapidity_i t.
    \label{equ:stationary_phase}
\end{align}
Here $\varphi(\rapidity) = \phi'(\rapidity)$. In Appendix \ref{app:Bethe_phase_concave} we show that there always exists a unique solution $\bs\theta = \bs{\theta}(t,\bs{x},\bs{y}^0)$ at least for time $t > t\ind{c} = \tfrac{N}{\ell}\sup_\rapidity\abs{\varphi'(\rapidity)}$. 

A necessary condition for the stationary phase approximation to make sense is that the functions of $\theta_i$ against which the fast-oscillating phase is integrated, are smooth enough, so that their variation scale is much larger than the oscillation scale of the phase. The oscillation scale is $1/\ell$, so the lower bound of the requirement \eqref{equ:widthmore} guarantees that this holds. We discuss at more length the validity of the stationary phase approximation in Sec.~\ref{sec:gas_discussion}.

With the upper bound of \eqref{equ:widthmore}, we can now take $\Delta \rapidity \to 0$. This indicates that the term in the wave function with trivial permutation, $\sigma = \bf 1$, {\em vanishes least quickly} if the coordinates stay along the trajectory $\bs x = \bs x(t)$ given by $\rapidity_i(t,\bs{x}(t),\bs{y}^0) \approx \rapidity_i^0$:
\beq
    \Psi(\ell t,\ell\bs x(t))|_{\sigma = \bf 1}\to 0 \quad \mbox{least quickly if} \quad
    y_i^0 + \rapidity_i^0 t = x_i(t) + \frac{1}{2\ell} \sum_{j\neq i}\sgn(x_i(t)-x_j(t)) \varphi(\rapidity_i^0-\rapidity_j^0).
    \label{equ:Psi_SP_sigma1}
\eeq
Solving this gives the solution  $x_i(t) = \b x_i(\bs y^0+\bs\theta^0 t,\bs\theta^0)$ in terms of functions $\b x_i(\bs y,\bs\theta)$ defined by solving
\beq\label{xbar}
    y_i = \b x_i(\bs y,\bs\theta) + \frac{1}{2\ell} \sum_{j\neq i}\sgn(\b x_i(\bs y,\bs\theta)-\b x_j(\bs y,\bs\theta)) \varphi(\rapidity_i-\rapidity_j).
\eeq
For the other terms in the sum over permutations, we get that the dominant part is at $x_i(t) = \b x_i(\bs y_{\sigma^{-1}}^0+\bs\theta_{\sigma^{-1}}^0 t,\bs\theta_{\sigma^{-1}}^0)$.

The equations $x_i(t) = \b x_i(\bs y^0+\bs\theta^0 t,\bs\theta^0)$ are the particle trajectories representing the wave packets approximate positions, at large scales, in macroscopic coordinates. Because of the sum over permutations, there is no unique set of particle trajectories, but it is the combined effect, in the large-scale limit, that should give the positions, as function of time, of the dominant part of the full wave function.  We discuss these  trajectories in Section \ref{sec:soliton_gas_interpretation}

We will discuss in the next section how this is connected to GHD. For now, we would like to note that after the stationary phase approximation on each permutation term, the wave function becomes
\begin{align}
	\Psi(\ell t,\ell\bs{x}) &\approx \frc{c_N s(\bs x)}{\sqrt{(2\pi \ell)^N}} \sum_{\sigma\in \mathcal{S}_N} (-1)^{|\sigma|} \frac{e^{\ri\ell S_t(\bs{x}_{\sigma},\bs{y}^0,\bs{\rapidity}(t,\bs{x}_{\sigma},\bs{y}^0))}}{\sqrt{\det \vb{H}(t,\bs{x}_{\sigma},\bs{y}^0,\bs{\rapidity}(t,\bs{x}_{\sigma},\bs{y}^0))}} \prod_i{A}_i\qty(\tfrac{\rapidity_i(t,\bs{x}_{\sigma},\bs{y}^0)-\rapidity^0_i}{\Delta \rapidity}).\label{equ:LL_SP_SP_WF}
\end{align}
Here, $\vb{H}(t,\bs{x},\bs{y}^0,\bs{\theta})$ is the Hessian of the Bethe phase, i.e $\vb{H}_{ij}(t,\bs{x},\bs{y}^0,\bs{\theta}) = \partial_{\rapidity_i}\partial_{\rapidity_j}S_t(\bs{x},\bs{y}^0,\bs{\rapidity})$. 

\subsection{Approximation Scheme 2: Linearisation of Bethe phase}\label{ssectapprox2}

We present another route to arrive at the same classical particle trajectories, which applies in a different regime. As we will explain in Sec.~\ref{sec:gas_discussion}, this works under a different restriction from the range \eqref{equ:width}:
\beq\label{equ:widthless}
    1/\ell \leq \Delta \rapidity \ll 1/{\sqrt\ell}.
\eeq

We first recall that the stationary-phase calculation of the Sec.~\ref{ssectapprox1} may fail for early macroscopic times $t<t\ind{c}$ (in the non-interacting limit $c=0$ of \eqref{equ:pre_int_LL_Nparticle}, it works for strictly positive times only, as otherwise the phase of the wave function is linear). Nevertheless, the effective trajectories in \eqref{equ:Psi_SP_sigma1} make sense at early times, and one expects hydrodynamic behaviours to emerge from infinitesimal macroscopic times. The alternative route we present works at all times, including \change{(macroscopic) $t=0^+$}. Intuitively, this is because, under \eqref{equ:widthless}, the extent of wave packets is large enough in real space so that there is enough ``mixing'' already from \change{$t=0^+$}.

Let us illustrate this for one particle, $N=1$. Consider the amplitude modulation \eqref{equ:amplitudemodulationfree}. According to \eqref{equ:LL_SP_psi_starting_point} (at $N=1$), the Fourier modes associated to it evolve simply as $A(\theta)\to A(\theta)e^{-\frc{\ri}2\theta^2 t}$. Therefore, in scaled coordinates, the amplitude modulation Eqs.~\eqref{equ:amplitudemodulationfree}, \eqref{equ:Atilde} evolves as
\beq\label{Psionepartevol}
    \Psi(\ell t,\ell x)
    =
    \frc{e^{-\ri (\ell\theta^0 y^0 +\frc{1}2(\theta^0)^2 \ell t)}}{\ell \sqrt{2\pi}}\t A(t,x-y_0)e^{\ri \ell\theta^0 x}
\eeq
where, recalling the relation $\alpha = \ell(\theta-\theta^0)$ from \eqref{equ:LL_SP_psi_amplitude_free} and \eqref{equ:Atilde} (with $\Delta\theta(\ell) = 1/\ell$),
\beq
    \t A(t,x) = 
    \int\dd{\alpha}
    A_1(\alpha)
    e^{\ri\alpha x-\frc{\ri}2 \big((
    \theta^0 + \ell^{-1}\alpha)^2
    -
    (
    \theta^0)^2\big)\ell t}
    =
    \t A(x-\theta^0 t) +\mathcal O(\ell^{-1}).
\eeq
Thus, apart from a physically irrelevant $x$-independent pure phase in \eqref{Psionepartevol}, the large-scale time evolution is, to leading order in $1/\ell$, given by translating the modulating function (recall the general relation \eqref{modulatingfctinteracting} between $B(\bs x)$ and $\t A(\bs x,\bs y^0)$, which holds also at finite times) as
\beq\label{Psionepartevol}
    \Psi(\ell t,\ell x)
    =
    \frc{e^{-\frc\ri2(\theta^0)^2 \ell t}}{\sqrt{2\pi}}B(t,x) e^{\ri \ell\theta^0 x}
\eeq
with
\beq\label{freemodtime}
    B(t,x+x(t)) = B(0,x),\quad  x(t) = y^0 + \theta^0 t.
\eeq
This is the non-interacting, $N=1$ version of the trajectories in \eqref{equ:Psi_SP_sigma1}.

We parallel this computation in the interacting case. \change{First, we take the wave packet example, Eq.~\eqref{modulatingfctinteracting}, of amplitude modulations.} The fundamental relation is the following, which expresses how the Bethe phase changes under small $\ep_i$ shifts of rapidities:
\beq\label{phibardisplacement}
    \b\Phi(\bs x,\bs\theta+\bs\ep)
    = \b\Phi(\bs x,\bs\theta)
    +
    \bs\ep\cdot \b{\bs y}(\bs x,\bs\theta) + \mathcal O(N\ep^2)
\eeq
where the functions $\b y_i(\bs x,\bs\theta)$ are the inverse of \eqref{xbar},
\beq\label{ybar}
    \b y_i(\bs x,\bs\theta) = x_i + \frac{1}{2\ell} \sum_{j\neq i}\sgn(x_i-x_j) \varphi(\rapidity_i-\rapidity_j).
\eeq
We will refer to the functions $\b y_i(\bs x,\bs\theta)$ as the {\em contracted macroscopic coordinates}. Note that the correction \change{$\mathcal O(N\ep^2)$} is absent in the free case. From \eqref{equ:LL_SP_psi_starting_point} and \eqref{equ:LL_SP_psi_amplitude} we see that the wave packet functions evolve in a simple way, with time $\ell t$:
\beq
    A_i(\alpha_i) \to A_i(\ell t,\alpha_i)
    = A_i(\alpha_i)e^{-\frc\ri2 (\theta_i^0 + \Delta\theta\,\alpha_i)^2 \ell t}.
\eeq
Accordingly, in the time-evolved modulation
\beq\label{Psimodulatedinteractingtime}
    \Psi(\ell t,\ell x) = 
    \t c_N(t) s(\bs x)\sum_{\sigma\in S_N}
    (-1)^{|\sigma|}
    \t A(t,\bs x_\sigma,\bs y^0)
    e^{\ri\ell\b \Phi(\bs x_\sigma,\bs\theta^0)
    }
\eeq
with $\t c_N(t) = e^{-\ri( \ell\bs\theta^0\cdot \bs y^0 +\frc12 (\bs \theta^0)^2\ell t)}(\Delta\theta)^Nc_N$, the modulating function evolves, using $\Delta\theta\to0$, as
\beq\label{modulatingfctinteractingtime}
    \t A(t,\bs x,\bs y^0) =
    \int \dd[N]{\alpha}
    \prod_i A_i(\alpha_i)
    e^{\ri \ell \Delta\theta \bs\alpha \cdot (\b{\bs y}(\bs x,\bs\theta^0) - \bs y^0 - \bs\theta^0 t) + \mathcal O(N\ell\Delta\theta^2)}.
\eeq
Neglecting the $O(N\ell\Delta\theta^2)$ term in the exponential (see the discussion in Subsec.~\ref{sec:gas_discussion}), we may perform the $\alpha_i$ integrals to get
\begin{align}\label{modulationevolve}
    \t A(t,\bs x,\bs y^0) \approx
    \prod_i \t A_i\big(\ell\Delta\theta
    (\b y_i(\bs x,\bs\theta^0)-y_i^0 - \theta_i^0 t)\big)
\end{align}
where $\t A_i(y) = \int\dd{\alpha}{A}_i(\alpha)e^{i\alpha y}$ is the Fourier transform of the rapidity-space wave packet ${A}_i(\alpha)$.

The result \eqref{modulationevolve} shows two things that happen in this approximation: (1) the modulating function for the gas of wave packets factorises as a product of the real-space wave packets $\t A_i(y)$ of the individual modes; and (2) these real-space wave packets are evaluated at linearly-displaced contracted macroscopic coordinates $\b y_i(\bs x,\bs\theta)$, with linear displacement from the initial coordinates $y_i^0$, and from the time evolution given by a velocity $\theta_i^0$. 

If we assume $\Delta\theta\gg 1/\ell$ and thus $\ell \Delta \theta_i \gg 1$, \change{the wave function \eqref{Psimodulatedinteractingtime} with \eqref{modulatingfctinteractingtime} has the structure that for every coordinate $\alpha_i$ over which there is an integration, there is a fast oscillating phase, unless coordinates are chosen appropriately such that this phase has vanishing coefficient. Note that the correction term $\mathcal O(N\ell\Delta\theta^2)$ is $\mathcal O(\ell\Delta\theta^2)\ll \mathcal O(\ell\Delta\theta)$ on each coordinate, so can be neglected in making this choice.} Therefore, the wave function is dominant for $x_i=x_i(t)$ with $\b y_i(\bs x_{\sigma^{-1}}(t),\bs\rapidity^0) = y_i^0+\theta_i^0 t$, which is the same condition as that for the trajectories $\bs x(t)$ in \eqref{equ:Psi_SP_sigma1} (and its coordinate permutations). We obtain, again, trajectories of interacting point particles $x_i(t) = \b x_i(\bs y^0+\bs\theta^0 t,\bs\theta^0)$ (and their permutations).

In the case $\Delta\theta=1/\ell$, \change{there is no fast oscillating phase, but the leading term in the exponential is $\mathcal O(N)$ because of the sum over $N$ particles, while the correction term is $\mathcal O(1)$. Hence, in a large-deviation sense, it is dominant (see Sec.~\ref{sec:gas_discussion} for a discussion).} Thus, in terms of the non-contracted macroscopic coordinate functions $\b x_i({\bs y},\bs\theta^0)$ as defined in \eqref{xbar}, we find an evolution equation for the modulating function that generalises \eqref{freemodtime} to a translation of contracted coordinates,
\beq\label{modulatingfctevol}
    B(t,\b{\bs x}({\bs y}+\bs\theta^0 t,\bs\theta^0),\bs y_0)
    \approx 
    B(0,\b{\bs x}({\bs y},\bs\theta^0),\bs y_0)
    \approx
    e^{-\ri \ell\bs\theta^0\cdot \bs y^0} \ell^{-N}\prod_i \t A_i\big( y_i- y_i^0\big), \quad
    \Delta\theta = 1/\ell
\eeq
where $\bs y$ plays the role of the independent variable for the spatial variation of the modulating function. Here, the wave function (for a given permutation) does not concentrate on fixed trajectories because real-space wave packets are of finite macroscopic extent. Nevertheless, the evolution of the amplitude is still controlled by the trajectories $\bs x(t)$ in \eqref{equ:Psi_SP_sigma1}. Because wave packets are of macroscopic extent, these displacements represent a slow, macroscopic-scale time evolution.

Our conclusion of coordinate displacements, for $\Delta\theta=1/\ell$, is a more general phenomenon, which {\em does not require the factorisation of the amplitude in terms of individual wave packets}. Take the amplitude modulation \eqref{Psimodulatedinteractinggeneral}, which is not necessarily a gas of wave packets. Although it is not of the form \eqref{equ:LL_SP_psi_starting_point} with \eqref{equ:LL_SP_psi_amplitude}, we may still evaluate its time evolution. This is done by what we refer to as the {\em Bethe-Fourier transform} of the modulating function:
\beq\label{BF}
	B(\bs x) = \int \dd[N]{\alpha} A(\bs \alpha) e^{\ri \bs\alpha\cdot\b{\bs y}(\bs x,\bs\theta)}
    \quad \Leftrightarrow\quad
    A(\bs \alpha) = \int \frc{\dd[N]{y}}{(2\pi) ^N} B(\b{\bs x}(\bs y,\bs\theta)) e^{-\ri \bs\alpha\cdot \bs y}.
\eeq
Inserting $B(\bs x)$ as written in \eqref{BF} into \eqref{Psimodulatedinteractinggeneral}, and using \eqref{phibardisplacement}, we find
\beq
    \Psi(0,\ell x) = \int \dd[N]{\alpha}
    A(\bs \alpha)\,
    c_Ns(\bs x)\sum_{\sigma\in S_N}
    (-1)^{|\sigma|}
    e^{\ri\ell\b \Phi(\bs x_\sigma,\bs\theta^0 + \bs\alpha/\ell) + \mathcal O(N/\ell)
    }
    =
    \int \dd[N]{\alpha}
    A(\bs \alpha)
    \psi(\bs x_\sigma|\bs\theta^0+\bs \alpha/\ell)
\eeq
which is expressed in terms of the eigenfunctions $\psi(\bs x_\sigma|\bs\theta^0+\bs \alpha/\ell)$, hence from which time evolution is immediate and gives
\beq\label{Psimodulatedinteractingtimegeneral}
    \Psi(\ell t,\ell x) = 
    c_N(t) s(\bs x)\sum_{\sigma\in S_N}
    (-1)^{|\sigma|}
    B(t,\bs x_\sigma)
    e^{\ri\ell\b \Phi(\bs x_\sigma,\bs\theta^0)
    }
\eeq
where $c_N(t) = e^{-\frc\ri2 (\bs \theta^0)^2\ell t}c_N$,
along with
\beq\label{modulatingfctevolgeneral}
    B (t,\b{\bs x}({\bs y}+\bs\theta^0 t,\bs\theta^0))
    \approx 
    B(\b{\bs x}({\bs y},\bs\theta^0)).
\eeq
Thus, we see that if the factorisation \eqref{equ:LL_SP_psi_amplitude} does not hold, quantum superposition of such gases of wave packets still give rise to the same particle trajectories, for coordinates in the corresponding slowly-varying amplitude modulation.


\begin{remark}
    We note that \eqref{equ:LL_SP_SP_WF}, \eqref{Psimodulatedinteractingtime} and \eqref{Psimodulatedinteractingtimegeneral} are concrete approximations of how we expect the microscopic wave function to look like on the hydrodynamic scale, i.e. the equivalent of the hydrodynamic approximation on the level of the wavefunction. Such expressions are an important new feature of the ab-initio derivation: the standard derivation of hydrodynamics based on thermodynamics cannot give insights into the coherent evolution of the wave function. We believe that such explicit approximations will be crucial in further understanding and eventually proving the emergence of GHD in quantum systems.
\end{remark}

\subsection{Soliton gas interpretation}\label{sec:soliton_gas_interpretation}

In both derivations above, we have identified a set of particle trajectories in macroscopic coordinates: using the functions \eqref{xbar},
\beq
    x_i(t) = \b{\bs x}(\bs y^0 + \bs\theta^0 t,\bs\theta^0).
\eeq
These trajectories control time evolution: either as positions of spatially sharp wave packets, or coordinates displacements of the the modulating function in the amplitude modulation. Dropping the upper-script 0 on the rapidity, these are
\begin{align}
	y_i^0 + \rapidity_i t &= x_i(t) + \frac{1}{2\ell} \sum_{i\neq j} \sgn(x_i(t)-x_j(t))\varphi(\rapidity_i-\rapidity_j).\label{equ:LL_SP_QP_def}
\end{align} 
and going back to microscopic coordinates, 
\begin{align}
    y_i^0 + \rapidity_i t &= x_i(t) + \frac{1}{2} \sum_{i\neq j} \sgn(x_i(t)-x_j(t))\varphi(\rapidity_i-\rapidity_j).\label{equ:LL_SP_QP_def_micro}
\end{align}
In the microscopic coordinates, instead of $\b{\bs y}(\cdot,\bs\theta)$ and $\b{\bs x}(\cdot,\bs\theta)$, we denote the corresponding maps $\bs y^{\bs\theta}$ and its inverse $\bs x^{\bs \theta}$, which are analysed in Appendix \ref{app:scattering_map}:
\beq\label{ythetamap}
    y_i^{\bs \theta}(\bs x) =
    x_i + \frac{1}{2} \sum_{i\neq j} \sgn(x_i-x_j)\varphi(\rapidity_i-\rapidity_j),\quad
    y_i^{\bs\theta}(\bs x^{\bs\theta}(\bs y))
    = y_i.
\eeq

For fixed $t, \rapidity_i$ and $y_i^0$ a solution to \eqref{equ:LL_SP_QP_def} does not necessarily exists because of the jump of the $\sgn(x_i(t)-x_j(t))$.  There are two simple ways to overcome this problem
\begin{itemize}
    \item The condition \eqref{equ:LL_SP_QP_def} can be seen as a minimization condition of the convex function
    \begin{align}
	\mathcal{A}_t(\bs{x},\bs{\rapidity},\bs{y}^0) &= \tfrac{1}{2} \sum_i \qty(x_i - y_i^0 - \rapidity_i t)^2 + \tfrac{1}{4\ell} \sum_{i\neq j} \abs{x_i-x_j}\varphi(\rapidity_i-\rapidity_j).\label{equ:LL_SP_QP_action}
\end{align}
The minimizer of this functional always exists and is unique.
\item Alternatively, one can smoothen out the sign function, e.g. $\sgn_\alpha(x) = \tanh(x/\alpha)$ for a small $\alpha$, which will then produce a unique solution: the map $\bs y^{\bs\theta}$ is a diffeomorphism $\R^N\to\R^N$.
\end{itemize}

\begin{figure}[!h]
	\centering
	\includegraphics{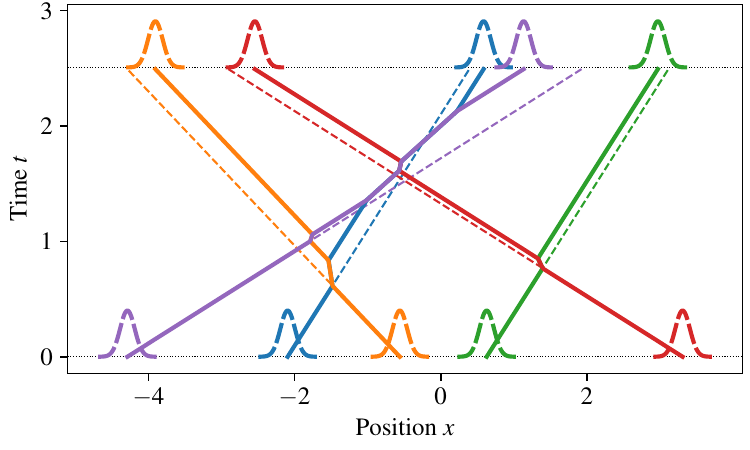}
	\caption{Evolution of Lieb-Liniger wave packets: the position of Lieb-Liniger wave packets (sketched as Gaussian bumps), follow trajectories (solid lines) given as solutions to \eqref{equ:LL_SP_QP_def}. During scattering particles `stick together' for a short amount of time, giving rise to an effective positions shift $\varphi(\rapidity-\rapidityp)$ compared to their non-interacting evolution (dashed lines).}
	\label{fig:LL_scbm}
\end{figure}

In figure \ref{fig:LL_scbm} we plot the trajectories of these particles obtained by minimizing \eqref{equ:LL_SP_QP_action}. We can see that far apart particles follow straight trajectories with the single particle velocity $v(\rapidity) = \rapidity$. During scattering, particles `stick together' for some time. Two particles scatter exactly for the amount of time such that after scattering their trajectories are shifted backwards by $\varphi(\rapidity_i-\rapidity_j)$. This can be thought as Wigner time delay occurring during the two-particle scattering. This is analysed in detail in Appendix \ref{app:scattering_map}. This behavior is analogous to the scattering of solitons in integrable PDEs~\cite{zabusky1965interaction}, and, on the large scale, gives rise to a soliton-gas phenomenology \cite{suret2023pre}. It agrees with the picture that generalized hydrodynamics describes a gas of interacting quasi-particles. Our derivation identifies these quasi-particles with wave packets of the quantum model.

If instead of defining the trajectories as minimizers of \eqref{equ:LL_SP_QP_action}, we smooth out the sign function, the trajectories in figure \ref{fig:LL_scbm} would also be smoothened. In this case, the particle trajectories \eqref{equ:LL_SP_QP_def} are those of a classical integrable model, studied in great detail in~\cite{PhysRevLett.132.251602,doyon2023generalisedtbartdeformationsclassicalfree}. In fact the analysis presented here was the motivation for the new integrable models defined in~\cite{PhysRevLett.132.251602,doyon2023generalisedtbartdeformationsclassicalfree}, which were called ``semi-classical Bethe models''.

In the hydrodynamic limit $\ell \to \infty$, \eqref{equ:LL_SP_QP_def} gives rise to generalized hydrodynamics. The quickest way to see this is to take a time-derivative of \eqref{equ:LL_SP_QP_def}
\begin{align}
    \rapidity_i &= \dot{x}_i(t) + \frac{1}{\ell} \sum_{i\neq j} \delta(x_i(t)-x_j(t))\varphi(\rapidity_i-\rapidity_j)(\dot{x}_i(t) - x_j(t))\label{equ:veff_deriv}
\end{align}
Now introduce $\rho(t,x,\rapidity) = \tfrac{1}{\ell} \sum_i \delta(x-x_i(t))\delta(\rapidity-\rapidity_i)$ (which will approach a continuous function as $\ell \to \infty$) and a velocity distribution $v\upd{eff}(t,x,\rapidity)$ defined via $v\upd{eff}(t,x_i(t),\rapidity_i) = \dot{x}_i(t)$. With this \eqref{equ:veff_deriv} becomes
\begin{align}
    v\upd{eff}(t,x,\rapidity) &= \rapidity + \int\dd{\rapidityp}\varphi(\rapidity-\rapidityp)\rho(t,x,\rapidityp) (v\upd{eff}(t,x,\rapidityp)-v\upd{eff}(t,x,\rapidity)),
\end{align}
and with $\p_t\rho(t,x,\theta) + \p_x \tfrac1\ell \sum_i \dot x_i(t)\delta(x-x_i(t))\delta(\theta-\theta_i)=0$, we obtain
\beq\label{ghdwavepackets}
	\p_t \rho(t,x,\theta,x) + \p_x \big(v^{\rm eff}(t,x,\theta)\rho(t,x,\theta)\big) = 0.
\eeq
This is precisely the GHD equation for the Lieb-Liniger model, Eq.~\eqref{ghd}. A more careful derivation can be found in~\cite{PhysRevLett.132.251602}, and in Subsec.~\ref{ssectheighfieldmethod} below.

\begin{remark}\label{rem:canonicaltrafo}
An alternative way to interpret \eqref{equ:LL_SP_QP_def} is to consider it as a map $\bs y^\theta$ from interacting coordinates $\bs x$ to non-interacting coordinates $\bs y$. This is what we call the ``Bethe scattering map''. Non-interacting coordinates $y_i(t) = y_i^0 + \rapidity_it$ evolve trivially. To solve the evolution one therefore has to (1) transform $\bs x\to \bs y$ at $t=0$, then (2) perform the trivial time evolution in $\bs y$ coordinates and (3) transform back to $\bs x$ coordinates at time $t$. The last step is the most tricky, in particular since the map $\bs x\to \bs y = \bs y^{\bs\theta}(\bs x)$ is not surjective. We discuss this in Appendix \ref{app:scattering_map}. Further, there is in fact a map $(\bs x,\bs p) \to (\bs y,\bs \theta)$ which generalises $\bs y^{\bs\theta}$ to include the physical momentum $\bs p$ \cite{doyon2023generalisedtbartdeformationsclassicalfree}. It can be seen as a canonical transformation of classical non-interacting particles with Hamiltonian $H(\bs{y},\bs{\theta})=\tfrac{1}{2}\sum_i \rapidity_i^2$ with generating function given by the Bethe phase \eqref{equ:pre_int_LL_phase}. This realization is crucial for showing that the trajectories \eqref{equ:LL_SP_QP_def_micro} correspond to a locally-interacting, integrable Hamiltonian model, as done in~\cite{doyon2023generalisedtbartdeformationsclassicalfree}.
\end{remark}

\subsection{Discussion of the applicability of the approximation schemes}\label{sec:gas_discussion}

The two approximation schemes of the wave function \eqref{equ:LL_SP_SP_WF}, and \eqref{Psimodulatedinteractingtime} and \eqref{Psimodulatedinteractingtimegeneral}, are different. Intuitively, they describe the same physics: the wave packets evolve along the trajectories of the classical particle model, and in the general case \eqref{Psimodulatedinteractingtimegeneral} with \eqref{modulatingfctevolgeneral}, the coordinates of the general modulating function evolve along these trajectories. Nevertheless, \eqref{equ:LL_SP_SP_WF} is expressed in terms of the amplitudes ${A}_i(\alpha)$ in rapidity space, while \eqref{Psimodulatedinteractingtime}, in terms of their Fourier transforms $\mathrm A_i(y)$ in position space, and more generally \eqref{Psimodulatedinteractingtimegeneral} in terms of the Bethe-Fourier transform \eqref{BF}. Also \eqref{equ:LL_SP_SP_WF} includes a determinant in order to provide the leading asymptotic for each permutation term.

The differences between these two approximation schemes is not surprising, as {\em they are valid in different regimes}.

\change{In approximation scheme 1,} in \eqref{equ:LL_SP_SP_WF} we perform a stationary phase approximation, which approximates the phase by its quadratic Taylor polynomial. This way the integral turns into a Gaussian that can be explicitly evaluated (this gives rise to the determinant in \eqref{equ:LL_SP_SP_WF}). However, for that it is crucial that the amplitude $A_i\qty(\tfrac{\theta_i-\theta_i^0}{\Delta \theta})$, which varies over a range $\Delta\theta$, be constant over the Gaussian extent, which is $\mathcal O(1/\sqrt{\ell})$. Specifically, the Gaussian integral is obtained by rescaling to the finite variables $\delta \rapidity_i = \sqrt\ell(\rapidity_i - \rapidity_i(t,\bs x_{\sigma^{-1}},\bs y^0))$, and the integral obtained with the resulting $A_i\qty(\tfrac{\delta\theta_i}{\sqrt{\ell}\Delta \theta})$ as an asymptotic expansion as $\ell\to\infty$ only if $\tfrac{1}{\sqrt{\ell}}\tfrac{1}{\Delta\theta} \ll 1$, or equivalently
\beq
    \Delta \theta \gg 1/\sqrt{\ell}.
\eeq
The wave packets are broad in momentum space, but thin in position space (in macroscopic coordinates). Note that this includes the case $\Delta\theta = \order{1}$, in which case wave packets have microscopic spatial extent. By the Gaussian integral, the corrections are, for every permutation term, of the form $\prod_{i=1}^N\big(1+\mathcal O\big(\frc1{\ell\Delta\theta^2}\big)\big)$. Thus, with $N\propto \ell$, the full error exponentiates to
\beq\label{error1}
    e^{\mathcal O(1/\Delta\theta^2)}
    \ll e^{\mathcal O(\ell)}.
\eeq
Thus the approximation of the wave function is valid in a ``large-deviation'' sense (see also remark \ref{rem:LD}).

\change{In approximation scheme 2,} in \eqref{Psimodulatedinteractingtime} we neglect terms of order $\order{N\ell \Delta \rapidity^2}$ in the phase, within the exponential (a similar analysis holds for \eqref{Psimodulatedinteractingtimegeneral} with $\Delta\theta = 1/\ell$). As this is to be compared with a term $\mathcal O(N\ell)$, the explicit phase in \eqref{Psimodulatedinteractingtime} (using the fact that $\b\Phi(\bs x_\sigma,\bs\theta^0)\propto N$),  a term $\mathcal O(N\ell\Delta\theta)$, the leading phase in \eqref{modulatingfctinteractingtime}, and a term $\mathcal O(N)$, coming from the product of factors $\prod_i A_i(\alpha_i)$, we must have $\Delta\theta\ll 1$ and $\ell\Delta\theta^2\ll 1$, thus
\beq
    \Delta \rapidity \ll 1/\sqrt{\ell}.
\eeq
Wave packets are thin in momentum space, but broad in position space. This also includes the case $\Delta\theta = \order{1/\ell}$, where wave packets have macroscopic spatial extend, representing coherence over a macroscopic scale. We emphasise that, with $N\propto \ell$, Eq.~\eqref{equ:macrolimit}, for every permutation term, the error is
\beq\label{error2}
    e^{\mathcal O(\ell^2\Delta\theta^2)}\ll e^{\mathcal O(\ell)}.
\eeq
Again, the result is valid in a ``large-deviation'' sense.

One aspect that we do not know how to properly account for is the sum over permutations. Both the stationary phase approximation scheme and the linearisation of the phase approximation scheme are asymptotic, to hold in a large-deviation sense as explained above. It is certainly not justified to simply take a sum over all of these $N!\sim \ell!$ terms, even more so as each has its own fast oscillating phase. Therefore, we do not expect that neither \eqref{equ:LL_SP_SP_WF} nor \eqref{Psimodulatedinteractingtime}, \eqref{Psimodulatedinteractingtimegeneral} are actual large-deviation forms for the wave function, neither in a pointwise nor in a $L^2$ sense.

Nevertheless, we believe these approximation schemes provide the correct intuitive understanding of what is going on on large scales in Bethe ansatz wave functions, and of how GHD emerges in integrable quantum systems.



\begin{remark}
    \change{Note that the special case $\Delta \rapidity = \order{1/\sqrt{\ell}}$, is excluded from both approximation schemes. In this case we can (for instance) still use \eqref{Psimodulatedinteractingtime}, but including the second order approximation. This leads to an integral over $\alpha$ that depends on the amplitude ${A}(\alpha)$.}
\end{remark}

\begin{remark}\label{rem:LD}
    \change{Why is it justified to neglect (potentially) divergent factors of order $\ll e^{\order{\ell}}$ when the leading factor is $e^{\mathcal O(\ell)}$?
    
    Consider the following example. Setting $0\leq \nu<1$, it is well known that, for well behaved functions $f(\alpha),\,g(\alpha)$ we may evaluate the following integral by a stationary phase approximation:
    \beq\label{examplesimple}
    \int \dd \alpha \,e^{\ri (\ell f(\alpha) + \ell^\nu g(\alpha))}
    =
    \sum_{\alpha^*:f'(\alpha^*)=0}e^{\ri \ell f(\alpha^*) + \mathcal O(\ell^\nu)}.
    \eeq
    We use a similar principle in our derivation, extending this to $N$ integration variables with $N\propto \ell$. The simultaneous limit $N\propto\ell\to\infty$ is well understood in the ``classical'' case, with real exponentials; this can be formalised within large-deviation theory, where the integral is dominated by the region that is  least exponentially small (for a use of such ideas in classical integrable models, see for instance~\cite[Sec. 4.4]{phdfriedrich}, and \cite[App. B]{doyon2023generalisedtbartdeformationsclassicalfree}). For oscillatory exponential, as in the simple example above, this corresponds to neglecting regions of integration with fast oscillations, and considering that the fastest-oscillating factor is what determines the leading region of integration. Such ``infinite dimensional'' stationary phase approximations are frequently used in the physics literature, a well-known example being the semi-classical limit of the Feynman path integral in quantum mechanics~\cite{feynman2005feynman}.

    Why is it sufficient, in our calculations of wave functions, to keep only the leading term in the exponent, as on the right-hand side of \eqref{examplesimple}? This is, intuitively, because the wave function is akin to a probability distribution: quantum averages will be supported on slowly oscillating regions.

}
\end{remark}

\section{Computing GHD of expectation values}\label{sec:expecation_values}

The last section shows that in the Lieb-Liniger model wave packets evolve like its GHD quasi-particles, and in the large scale limit the density of wave packets satisfies the GHD equation \eqref{ghdwavepackets}. While this might seem expected, it is actually very surprising. The density of wave-packets is not the quasi-particle density of GHD. The physical quasi-particle density is defined via expectation values of its local charge densities. For instance, due to the emergent Pauli exclusion principle in the Lieb-Liniger model -- the effect of the anti-symmetric sum over permutations in \eqref{equ:pre_int_LL_Nparticle} -- the density of wave packets is not simply related to the observed density, in particular in high density regions. The fact that both the density of wave packets (as we have shown above, Eq.~\eqref{ghdwavepackets}) and the physical densities (as expected, Eq.~\eqref{ghd}) satisfy the GHD equation, is rather mysterious.

In this section, we propose potential ways to obtain GHD equation for the physical densities.

We start with a precise result, the explicit construction of physical densities, including the spectral phase-space density operator, which is a crucial object in GHD.

Then, we consider two ways of deriving the GHD equation for expectation values of this operator. First, we use our approximations \eqref{equ:LL_SP_SP_WF} and \eqref{Psimodulatedinteractingtime}, \eqref{Psimodulatedinteractingtimegeneral}, both for the gas of wave packets \eqref{equ:LL_SP_psi_starting_point}, and more physical locally entropy-maximised states, which we build out of amplitude modulations \eqref{Psimodulatedinteractingtimegeneral}. Second, we re-work the large-scale analysis of the expectation value directly, using an appropriate Wigner quasi-probability distribution -- this is our third proposal to obtain GHD.

These derivations of the GHD equation, for expectation values of the spectral phase-space density operator, are, currently, incomplete. We point out the flaws and potential completions.

\subsection{Action of local conserved densities on Bethe states}

In order to make sense of $\rho\ind{p}(t,x,\rapidity)$ as an expectation value, we need to define a corresponding quantum operator $\quasiparticleop(x,\rapidity)$. We will define this operator explicitly in the following way on a Bethe state
\begin{align}
    \quasiparticleop(x,\rapidity)\psi(\bs{x}|\bs{\rapidity}) &= c_N s(\bs{x})\sum_{\sigma\in \mathcal{S}_N} (-1)^{|\sigma|} \qty[\sum_i \delta(x-x_{\sigma_i})\delta(\rapidity-\rapidity_i)] e^{\ri\Phi(\bs{x}_{\sigma},\bs{\rapidity})}.\label{equ:LL_deriv_cq_def}
\end{align}
This, since Bethe states are a basis of $L^2\ind{sym}(\mathbb{R}^N)$, completely characterizes the operator. This definition was chosen in such a way that $\hat{q}_n(x) = \int\dd{\rapidity}\quasiparticleop(x,\rapidity)\rapidity^n$ is a local density of the $n$'th charge $\hat{Q}_n$ in the Lieb-Liniger model. Indeed, one can easily check:
\begin{align}
    \hat{Q}_n\psi(\bs{x}|\bs{\rapidity}) &= \int\dd{x}\hat{q}_n(x)\psi(\bs{x}|\bs{\rapidity}) = \int\dd{x}\dd{\rapidity}\rapidity^n\quasiparticleop(x,\rapidity)\psi(\bs{x}|\bs{\rapidity}) = \qty[\sum_i \rapidity_i^n]\psi(\bs{x}|\bs{\rapidity}),
\end{align}
which is exactly the action of the $n$'th charge on a Bethe state. It is not so straight-forward, however, to establish that these densities are indeed local. We show this in Appendix \ref{app:densities_of_cqs}. \change{We would like to note that the conserved densities of the Lieb-Liniger model have been extensively studied in the literature and explicitly constructed; see in particular \cite{davieshigher,davieskorepin} and also the perturbative constructions~\cite{PhysRevE.98.052126,PhysRevLett.128.190401}. However, the action of these operators on Bethe states is not explicitly known. Therefore, we would like to stress that the explicit action of charge densities on Bethe states as given by \eqref{equ:LL_deriv_cq_def} is a major result of this work. }

\begin{remark} The operator $\quasiparticleop(x,\rapidity)$ (and thus also the $\hat{q}_n(x)$) defined via \eqref{equ:LL_deriv_cq_def} is not necessarily hermitian. This can easily be fixed by defining $\quasiparticleop(x,p) \to \tfrac{1}{2}(\quasiparticleop(x,p) + \quasiparticleop(x,p)^\dagger)$. This replacement, however, will not affect the following computations. Hence, for sake of simplicity, we will stick with \eqref{equ:LL_deriv_cq_def}.
\end{remark}

\subsection{Wave function approximation for the gas of wave packets}\label{ssectghdwavefunction}

We start with the gas of wave packets \eqref{equ:LL_SP_psi_starting_point}, and ask about the expectation value
\beq\label{equ:expval}
    \rho\ind{p}(t,x,\theta)
    =
    \frc{\langle{\Psi(\ell t)}|
    \quasiparticleop(\ell x,\rapidity)|
    {\Psi(\ell t)} \rangle}{
    \langle{\Psi(\ell t)}|
    {\Psi(\ell t)} \rangle
    }.
\eeq
We consider the cases where wave packets are point-like particles in the large-scale limit, and simultaneously, concentrate to point-like distribution in rapidity space as well: $1/\ell\ll \Delta\theta \ll 1$. As we will see, the main reason why computing \eqref{equ:expval} remains non-trivial even after making the large-scale wave function approximation, is because of the remaining sum over permutations.

\change{The first strategy to compute the expectation value is based on the stationary phase approximation \eqref{equ:LL_SP_SP_WF}, which in particular requires $1/\sqrt{\ell}\ll \Delta\theta\ll 1$. As discussed in Sec. \ref{sec:gas_discussion} the precise validity of this approximation is not clear, but let us proceed anyway and see what we find.} In order to use this approximation, we need to apply it also to the result of $\hat\rho(\ell x,\theta)|\Psi(\ell t)\rangle$, using \eqref{equ:LL_deriv_cq_def}, instead of just the wave function $|\Psi(\ell t)\rangle$. The stationary phase analysis of Sec.~\ref{ssectapprox1} goes through unchanged, if we integrate against well-behaved functions $\int \dd{x}\dd{\theta}f(x,\theta) \hat\rho(\ell x,\theta)|\Psi(\ell t)\rangle$. Thus, using \eqref{equ:LL_SP_SP_WF} and its simple modification for $\hat\rho(\ell x,\theta)|\Psi(\ell t)\rangle$, we obtain the following result, in a distributional sense on $x,\theta$:
\beq\begin{aligned}
	\rho\ind{p}(t,x,\theta) &\propto \int\dd[N]{x} \sum_{\sigma,\tau \in \mathcal{S}_N}(-1)^{|\sigma|} (-1)^{|\tau|}\sqrt{\frac{1}{\det \vb{H}(t,\bs{x}_{\sigma},\bs{y}^0,\bs{\rapidity}(t,\bs{x}_{\sigma},\bs{y}^0))}\frac{1}{\det \vb{H}(t,\bs{x}_{\tau},\bs{y}^0,\bs{\rapidity}(t,\bs{x}_{\tau},\bs{y}^0))}}\\
    &\quad \times \prod_i {A}_i\qty(\tfrac{\rapidity_i(t,\bs{x}_{\tau},\bs{y}^0)-\rapidity^0_i}{\Delta \rapidity})^*{A}_i\qty(\tfrac{\rapidity_i(t,\bs{x}_{\sigma},\bs{y}^0)-\rapidity^0_i}{\Delta \rapidity})\qty[\tfrac{1}{\ell}\sum_i \delta(x-x_{\sigma_i})\delta(\rapidity-\rapidity_i(t,\bs{x}_{\sigma},\bs{y}^0))]\\
    &\quad \times \exp[\ri\ell \qty(S_t(\bs{x}_{\sigma},\bs{y}^0,\bs{\rapidity}(t,\bs{x}_{\sigma},\bs{y}^0))-S_t(\bs{x}_{\tau},\bs{y}^0,\bs{\rapidity}(t,\bs{x}_{\tau},\bs{y}^0)))].
\end{aligned}\eeq
Observe that we still have a fast oscillating phase left. \change{At this point we perform another approximation: assuming that the fast oscillating phases cancel out}, the only configuration where no cancellation can occur is when $\sigma = \tau$. Thus, under the assumption that we can neglect all the terms where $\sigma \neq \tau$, we arrive at
\beqa
    \lefteqn{\rho\ind{p}(t,x,\theta)} \\  &\propto& \int\dd[N]{x} \frac{1}{\det \vb{H}(t,\bs{x},\bs{y}^0,\bs{\rapidity}(t,\bs{x},\bs{y}^0))} \prod_i \abs{{A}_i\qty(\tfrac{\rapidity_i(t,\bs{x},\bs{y}^0)-\rapidity^0_i}{\Delta \rapidity})}^2\qty[\tfrac{1}{\ell}\sum_i \delta(x-x_i)\delta(\rapidity-\rapidity_i(t,\bs{x},\bs{y}^0))].\no
\eeqa

Here we removed the remaining permutation $\sigma$ by changing variable $\bs{x} \to\bs{x}_\sigma$. Now sending $\Delta \rapidity \to 0$, we find that $\rapidity_i(t,\bs x,\bs y^0) \to \rapidity_i^0$, meaning that trajectories will concentrate on $x_i = x_i(t) = \b{x}_i(\bs{y}^0+\bs\theta t,\bs{\rapidity}^0)$, Eq.~\eqref{xbar}. Thus, all $x_i$ integrals are killed, and the factors before the large square bracket are replaced by a fixed function of $t,\bs y^0,\bs\theta^0$, independent of $x,\theta$. By normalisation, this function must in fact be trivial, and we obtain
\begin{align}
    \rho\ind{p}(t,x,\theta) \to \tfrac{1}{\ell} \sum_i \delta(x-\b{x}_i(\bs{y}^0+\bs\theta t,\bs{\rapidity}^0))\delta(\rapidity-\rapidity_i^0).\label{equ:simple_deriv_final}
\end{align}
That is, $\rho\ind{p}(t,x,\theta) \to \rho(t,x,\theta)$, the wave packet density, which satisfies the GHD equation \eqref{ghdwavepackets}.

A similar calculation can be done by using \eqref{Psimodulatedinteractingtime} with \eqref{modulationevolve} and $1/\ell\ll \Delta\theta\ll 1/\sqrt\ell$. Again we have to work out, within this approximation scheme, the quantity $\int \dd{x}\dd{\theta}f(x,\theta) \hat\rho(\ell x,\theta)|\Psi(\ell t)\rangle$, and this is done by putting the additional factor $\sum_j f(x_j,\theta_j^0+\Delta \theta \alpha_j)$ inside the integral in \eqref{modulatingfctinteractingtime}. Because $\Delta\theta\to 0$, this amounts to the factor $\sum_j f(x_j,\theta_j^0)$ on the right-hand side of \eqref{modulationevolve}, which we then replace by the distribution $\sum_j \delta(x-x_j)\delta(\theta-\theta_j^0)$ for evaluating $\rho\ind{p}(t,x,\theta)$. The arguments above go through: the phases from \eqref{Psimodulatedinteractingtime} impose equality of the permutation elements $\sigma,\tau$, and the factors in \eqref{modulationevolve} impose $\b y_i(\bs x,\bs\theta^0) = y_0^0 + \theta_i^0 t$, giving again \eqref{equ:simple_deriv_final}.

These derivations are quick and show how both the stationary phase and Bethe linearisation approximation schemes carry over to the expectation values. However, we believe that these derivations are too naive. Here are our remarks.
\begin{itemize}
    \item In our construction of the gas of wave packets, we are free to choose the $y_i^0$'s and the $\rapidity_i^0$'s arbitrarily. But then, $\rho\ind{p}(t,x,\theta) \to \rho(t,x,\theta)$ contradicts the fermionic statistics of the quasi-particles. For instance, we are able to pack as many wave packet positions $\b x_i(\bs y^0,\bs\theta^0)$ as we want in any region of space, while this cannot happen for the physical quasi-particles. The difference comes from the anti-symmetric sum over permutations in \eqref{equ:LL_SP_psi_starting_point}.
    Thus $\rho\ind{p}(t,x,\theta)$ cannot be identified with $\rho(t,x,\theta)$. 
    \item In what step is this subtlety hidden? On the one hand, recall that the \change{approximation schemes can only be correct} in some large-deviation sense, because of errors of the form \eqref{error1}, \eqref{error2}. However, because $\hat\rho(\ell x,\theta)$ is, as a distribution, a large-scale operator, we believe the leading approximation is sufficient. On the other hand, the stationary phase approximation replaces the original integral by a Gaussian integral of width $\delta \rapidity_i \sim 1/\sqrt{\ell}$. While for each individual permutation this stationary phase approximation is justified, there is a huge number of permutation exchanging nearby particles, within the range $\sqrt{\ell}$ in rapidity space. Similarly, the linearisation of the Bethe phase holds for real-space wave packets of sizes $\gg \sqrt{\ell}$, and there is again a huge number of permutations exchanging nearby particles within that range. Therefore, we believe simply dropping all permutations except $\sigma=\tau$ is too naive.
\end{itemize}

\subsection{Wave function approximation for the local-equilibrium state}\label{ssectghdlocalequil}

The gas of wave packets \eqref{equ:LL_SP_psi_starting_point} is not a state that is typically realised in experiments. It is likely to encode the correct physics, however it is useful to consider more physically likely states. A natural set of states are the local-equilibrium state, usually written as generalised Gibbs ensembles (GGE) with slowly-varying  ``temperatures". Using the hermitian part of the spectral phase-space density operator, these states are represented by the (un-normalised) density matrix
\beq
    \h\rho = e^{-\frc\ell2\int \dd{x}\dd{\theta} \beta(x,\theta)\big(\h\rho(\ell x,\theta) + \h\rho(\ell x,\theta)^\dag\big)}
\eeq
for spectral ``generalised temperature'' $\beta(x,\theta)$. For every macroscopic position $x$, the function $\beta(x,\cdot)$ characterises the GGE perceived by local observables at $x$. We will see that the expectation of the time-evolved spectral phase-space density operator in this state can be written in terms of the {\em general amplitude modulation} approximation, \eqref{Psimodulatedinteractingtimegeneral}, in which, as we said, we take $\Delta\theta=1/\ell$. Hence, this is not directly representable as a gas of wave packets, yet our large-scale evolution theory can still be applied.

In Sec.~\ref{sec:heisenberg}, we study the operator formalism. We find in particular that extensive observables such as $\ell \int \dd{x}\dd{\theta} \beta(x,\theta)\h\rho(\ell x,\theta)$ have commutators that are extensively supported by with sub-extensive strength -- that is, still integrals of local observables over $\sim \ell$ regions, but with densities of strength $\mathcal O(\ell^{-1})$. See Eq.~\eqref{rhorhodag}. In particular, this means that we can write
\beq
    \h\rho
    =
    \hat \varrho^\dag e^{\int \dd x\,\h f(x) u(\ell x)} \h \varrho
\eeq
for some function $f(x)$ and local observable $\h u(x)$, with
\beq
    \hat\varrho = 
    e^{-\frc\ell 2\int \dd{x}\dd{\theta}\beta(x,\theta)\h\rho(\ell x,\theta)}.
\eeq

This large-scale commutation property also suggests that we can directly exponentiate the form \eqref{equ:LL_deriv_cq_def} of the action of $\h\rho(x,\theta)$ on the Bethe eigenfunctions -- although we have not performed a full analysis of this. Assuming this, we find, in macroscopic coordinates,
\beq
    (\h\varrho e^{-\ri \h H \ell t}\psi(\cdot|\bs\theta))(\ell\bs x)
    =c_N s(\bs x)
    \sum_{\sigma\in S_N}(-1)^{|\sigma|}
    e^{-\frc12
    \sum_i\beta(x_{\sigma(i)},\theta_i)
    +\ri\ell\b\Phi(\bs x_\sigma,\bs\theta)-\frc\ri2 \ell \bs\theta^2 t
    +\mathcal O(1)}
\eeq
which is (up to the phase $e^{-\frc\ri2\ell\bs\theta^2t}$) the amplitude modulation \eqref{Psimodulatedinteractinggeneral} with $\bs\theta^0 = \bs\theta$ and modulating function
\beq
    \mathrm A(\bs x) = e^{-\frc12
    \sum_i\beta(x_i,\theta_i)+\mathcal O(1)
    }.
\eeq
Note how {\em this is not} of the form \eqref{modulationevolve} at $t=0$, because it factorises in the $\bs x$ coordinates, not in $\b{\bs y}(\bs x,\bs\theta)$. Thus, this is the general form \eqref{Psimodulatedinteractinggeneral}, which evolves as per \eqref{Psimodulatedinteractingtimegeneral} with \eqref{modulatingfctevolgeneral}.

Writing the partition function as a trace in the basis of Bethe eigenfunctions,
\beq
    Z_t = \Tr\h\rho(\ell t)
    =
    \frc1{N!}\int \dd[N]{x}\dd[N]{\theta}
    \psi(\bs x|\bs\theta)^\dag
    e^{\ri \h H \ell t}\,\h\rho\, e^{-\ri \h H \ell t}
    \psi(\bs x|\bs\theta),
\eeq
we must evolve the above amplitude modulation to time $-t$, and we obtain
\beq\label{Zt}
    Z_t = \frc{c_N^2}{N!}\int \dd[N]{x}\dd[N]{\theta}
    \sum_{\sigma,\tau\in S(N)}
    (-1)^{|\sigma|+|\tau|}
    \mathrm A(\b {\bs x}_\sigma(\b{\bs y}(\bs x,\bs\theta) + \bs \theta t,\bs\theta))\,
    e^{\ri\ell(\b\Phi(\bs x_\sigma,\bs\theta)-\b\Phi(\bs x_\tau,\bs\theta))+\mathcal O(1)}
    .
\eeq
The average density is obtained from
\beq
    \rho\ind{p}(t,x,\theta)
    = 
    -\frc{\delta \log Z_t}{\delta\beta(x,\theta)}
    .
\eeq
Note how time evolution is simply given by the classical particle system \eqref{equ:LL_SP_QP_def} (in macroscopic coordinates) for the arguments of the modulating function.

Under the same assumption we made in Sec.~\ref{ssectghdwavefunction}, that the highly oscillatory phases will force $\sigma=\tau$, we get the time-evolved classical partition function for the system of particles \eqref{equ:LL_SP_QP_def_micro} in its local-equilibrium state. This partition function was analysed in \cite{doyon2023generalisedtbartdeformationsclassicalfree}. Within this partition function, a large-deviation analysis shows that the integration variables $(\bs x,\bs\theta)$ localise on a distribution $\rho(t,x,\theta)$, which therefore evolves according to \eqref{equ:simple_deriv_final} (for a choice of $\bs y^0,\,\bs\theta^0$ representing this distribution). Clearly, the same problem coming from the assumption that $\sigma=\tau$ occur here, as discussed in Sec.~\ref{ssectghdwavefunction}.

However, more generally, without making the assumption $\sigma=\tau$, it is likely that by a large-deviation analysis of \eqref{Zt}, the integration variables $(\bs x,\bs\theta)$ localise on a distribution $\rho(t,x,\theta)$ that encodes the Fermionic nature of the quasi-particles. Then, the time evolution would still be given by \eqref{equ:simple_deriv_final}, again for a choice of $\bs y^0,\,\bs\theta^0$ representing this distribution. An analysis of this is left for future works.

\subsection{Wigner quasi-probability distribution}\label{secwig}
In Sec.~\ref{ssectghdwavefunction} and \ref{ssectghdlocalequil} it became apparent that the sum over permutations (or equivalently the fermionic nature of the quasi-particles) was not taken into account properly. Perhaps the simplest way of taking it into account is to start from a fully antisymmetric amplitude $A(\bs{\rapidity})$. However, this is not so trivial, since for GHD we need two pieces of information per particle, its position $x_i$ and its rapidity $\rapidity_i$. An amplitude $A(\bs{\rapidity})$ is naturally interpreted as a distribution of the $\rapidity_i$, but the information about the positions $x_i$ is hidden in fast oscillating phases of the amplitude. To extract the position information as well, we choose to describe the state via its $N$ particle Wigner quasi-probability distribution~\cite{VITatarskii_1983}
\begin{align}
	W_N(\bs{y}^0,\bs{\rapidity}) &= \tfrac{N!}{\pi^N} \int\dd[N]{\rapidity'}A(\bs{\rapidity}-\bs{\rapidity}')A(\bs{\rapidity}+\bs{\rapidity}')^* e^{-2i\bs{\rapidity}'\bs{y}^0}.\label{equ:LL_deriv_2_wigner_def}
\end{align}
This function has the features of a phase-space distribution. In particular, its marginals are given by 
\begin{align}
	\int\dd[N]{y^0}W_N(\bs{y}^0,\bs{\rapidity}) &= N!\abs{A(\bs{\rapidity})}^2, & 
	\int\dd[N]{\rapidity}W_N(\bs{y}^0,\bs{\rapidity}) &= \frac{N!}{(2\pi)^N} \abs{\t {A}(\bs{y}^0)}^2
\label{equ:LL_deriv_2_wigner_marginals}
\end{align}
where $\t {A}(\bs{y}) = \int \dd[N]{\theta} A(\bs \theta) e^{\ri \bs\theta\cdot\bs y}$.

Note that the Wigner distribution can become negative in regions of size $\order{1}$. If we choose $A(\bs{\rapidity})$ to be a fully antisymmetric amplitude in \eqref{equ:LL_SP_psi_starting_point}, and appropriately normalised,
\begin{equation}
    \int \dd[N]{\theta} |A(\bs\theta)|^2 = \frc1{N!},
\end{equation}
then the expectation value \eqref{equ:expval} can be written as
\begin{align}
	\rho\ind{p}(t,x,\theta) &= \frac{\ell^N}{(2\pi)^N}\int\dd[N]{x}\dd[N]{y^0}\dd[N]{\rapidity}\dd[N]{\rapidity'} W_N(\bs{y}^0,\tfrac{\bs{\rapidity}+\bs{\rapidity}'}{2})e^{i\ell S_{t}(\bs{x},\bs{y}^0/\ell,\bs{\rapidity})-i\ell S_{t}(\bs{x},\bs{y}^0/\ell,\bs{\rapidity}')}\times\n
    & \qquad \times \qty[\tfrac{1}{\ell}\sum_i \delta(x-x_i)\delta(\rapidity-\rapidity_i)]\label{equ:LL_deriv_2_obs_basic}
\end{align}
where we use \eqref{equ:LL_SP_phase}.

In fact at this point, we can forget about the fact that the initial state is pure. If the initial state is mixed, then we can simply use \eqref{equ:LL_deriv_2_obs_basic} with $W_N(\bs{y}^0,\bs{\rapidity})$ being the Wigner quasi-probability distribution of that mixed state.

\begin{remark}
The Wigner distribution \eqref{equ:LL_deriv_2_wigner_def} is not the Wigner distribution of the physical particle. Instead, it is the Wigner distribution free fermions. We can interpret the Bethe wave functions as matrix elements of an unitary operator $\bra{\bs{x}}\hat{U}\ket{\bs{\rapidity}} = \psi(\bs{x}|\bs{\rapidity})$ mapping free fermionic wave functions onto Lieb-Linger wave functions. The mapping to free particles is a common technique in integrable systems, and also known as ``contracted coordinates'' in the context of GHD. These contracted coordinates are precisely the $y_i$, that evolve like non-interacting particles (see remark \ref{rem:canonicaltrafo}).
\end{remark}

Since in GHD we want to describe large scale dynamics, we will consider an $\ell$-dependent, slowly varying large scale Wigner distributions, which we normalise in order to simplify our equations:
\begin{align}
	W_{N}(\bs{y}^0,\bs{\rapidity}) = \frac{1}{\ell^{N}} \b W_{N}(\bs{y}^0/\ell,\bs{\rapidity})\label{equ:LL_deriv_2_wigner_scaling}
\end{align}
where $\b W_{N}(\bs{y}^0,\bs{\rapidity})$ is independent of $\ell$. Inserting this into \eqref{equ:LL_deriv_2_obs_basic}, we find
\begin{align}
    \rho\ind{p}(t,x,\theta) &= \frac{\ell^N}{(2\pi)^N}\int\dd[N]{x}\dd[N]{y^0}\dd[N]{\rapidity}\dd[N]{\rapidity'} \b W_{N}(\bs{y}^0,\tfrac{\bs{\rapidity}+\bs{\rapidity}'}{2})e^{i\ell S_{t}(\bs{x},\bs{y}^0,\bs{\rapidity})-i\ell S_{t}(\bs{x},\bs{y}^0,\bs{\rapidity}')}\qty[\tfrac{1}{\ell}\sum_i \delta(x-x_i)\delta(\rapidity-\rapidity_i)].
\end{align}
\change{In order to prepare for a large scale approximation}, let us denote the Fourier transform $\b{\mathrm{W}}_{N}(\bs{\rapidity},\bs{\rapidity}') = \int\dd[N]{y^0}\b{W}_{N}(y^0,\bs{\rapidity}')e^{-i\bs{\rapidity}\bs{y}^0}$ and use it to write
\begin{align}
	 \rho\ind{p}(t,x,\theta) &= \frac{\ell^N}{(2\pi)^N}\int\dd[N]{x}\dd[N]{\rapidity}\dd[N]{\rapidity'} \b{\mathrm{W}}_{N}(\ell(\bs{\rapidity}'-\bs{\rapidity}),\tfrac{\bs{\rapidity}+\bs{\rapidity}'}{2})e^{i\ell \bar{S}_t(\bs{x},\bs{\rapidity}) - i\ell \bar{S}_{t}(\bs{x},\bs{\rapidity}')}\qty[\tfrac{1}{\ell}\sum_i \delta(x-x_i)\delta(\rapidity-\rapidity_i)]
     \label{equ:LL_deriv_2_deriv_1}
    \\
	&=\frac{1}{(2\pi)^N}\int\dd[N]{x}\dd[N]{\rapidity}\dd[N]{q} \b{\mathrm{W}}_{N,\ell}(\bs{q},\bs{\rapidity}+\tfrac{1}{2\ell}\bs{q})e^{i\ell\bar{S}_t(\bs{x},\bs{\rapidity}) - i\ell\bar{S}_t(\bs{x},\bs{\rapidity}+\bs{q}/\ell)}\qty[\tfrac{1}{\ell}\sum_i \delta(x-x_i)\delta(\rapidity-\rapidity_i)].\label{equ:LL_deriv_2_deriv_2}
\end{align}
\change{So far our manipulations were exact. At this point we will take the $\ell\to\infty$ limit and keep the dominant terms. Since we also take $N\to \infty$ simultaneously, this is a completely uncontrolled approximation. Nonetheless, let us proceed and observe what we find:}
\begin{align}
	\rho\ind{p}(t,x,\theta) &= \frac{1}{(2\pi)^N}\int\dd[N]{x}\dd[N]{\rapidity}\dd[N]{q} \b{\mathrm{W}}_{N}(\bs{q},\bs{\rapidity}+\order{1/\ell})e^{-i\grad_{\rapidity}\bar{S}_{t}(\bs{x},\bs{\rapidity})\cdot\bs{q} + \order{N/\ell}}\qty[\tfrac{1}{\ell}\sum_i \delta(x-x_i)\delta(\rapidity-\rapidity_i)]\label{equ:LL_deriv_2_deriv_3}\\
	&= \int\dd[N]{x}\dd[N]{\rapidity} \b{W}_{N}(\b {\bs y}(\bs x,\bs\theta) - \bs \theta t),\bs{\rapidity})e^{\mathcal O(1)}\qty[\tfrac{1}{\ell}\sum_i \delta(x-x_i)\delta(\rapidity-\rapidity_i)]\label{equ:LL_deriv_2_deriv_4}.
\end{align}
Here,  $\bs{\rapidity}'=\bs{\rapidity}+\bs{q}/\ell$, we used \eqref{ybar} and we define $\b S_t(\bs{x},\bs{\rapidity}) = S_t(\bs{x},\bs y^0,\bs{\rapidity})+\bs\theta\cdot \bs y^0 \ell$,
\begin{align}
	\b S_t(\bs{x},\bs{\rapidity}) &= \sum_i \rapidity_i x_i +\frac{1}{4\ell}\sum_{i\neq j}\sgn(x_i-x_j) \phi(\rapidity_i-\rapidity_j)- \frac{1}{2}\sum_i\rapidity_i^2t.
\end{align}
Now, let us do a change of coordinates $\bs x \to \bs y^0 = \b{\bs y}(\bs x,\bs\theta)$. This map is not surjective but, except at jumps, it is linear of slope 1 (see Appendix \ref{app:scattering_map}). Therefore, we find
\begin{align}
    \rho\ind{p}(t,x,\theta)
    &=\int_{\Sigma_t}\dd[N]{y^0}\dd[N]{\rapidity} \b{W}_{N}(\bs{y}^0-\bs\theta t,\bs{\rapidity})e^{\mathcal O(1)}\qty[\tfrac{1}{\ell}\sum_i \delta(x-\b x_i(\bs{y}^0,\bs{\rapidity}))\delta(\rapidity-\rapidity_i)]
    \label{equ:LL_deriv_2_final_hatx}
\end{align}
where $\Sigma_t\subset \R^2$ is the image of $(\bs{x},\bs{\rapidity}) \mapsto \b{\bs{y}}(\bs{x},\bs{\rapidity})$.

\change{To understand the meaning of \eqref{equ:LL_deriv_2_final_hatx}, let us observe what happens if we neglect the missing space in $\Sigma_t$ and the correction terms $e^{\mathcal O(1)}$. We find}
\begin{align}
    \rho\ind{p}(t,x,\theta) &= \int\dd[N]{y^0}\dd[N]{\rapidity} \b{W}_{N}(\bs{y}^0,\bs{\rapidity})\qty[\tfrac{1}{\ell}\sum_i \delta(x-\b x_i(\bs{y}^0+\bs\theta t,\bs{\rapidity}))\delta(\rapidity-\rapidity_i)],\label{equ:wigner_no_det}
\end{align}
which is the evolution of particles along the trajectories $\b{x}_i(\bs{y}^0+\bs\theta t,\bs{\rapidity})$, whose initial data is distributed by $\b{W}_{N}(\bs{y}^0,\bs{\rapidity})$. Thus, ignoring that $\b{W}_{N}(\bs{y}^0,\bs{\rapidity})$ can be negative, \eqref{equ:wigner_no_det} is precisely the expression for $\rho\ind{p}(t,x,\theta)$ we would expect for the classical particle model \eqref{equ:LL_SP_QP_def_micro}.
\change{Furthermore, expression \eqref{equ:wigner_no_det} conserves the total particle number:}
\beq\label{rhopnorm}
    \int \dd{x}\dd{\theta} \rho\ind{p}(t,x,\theta) = N/\ell.
\eeq

\change{But instead of \eqref{equ:wigner_no_det} we find \eqref{equ:LL_deriv_2_final_hatx}}. Naively, the $e^{\mathcal O(1)}$ correction should not matter if $\b{W}_{N}(\bs{y}^0,\bs{\rapidity})$ is taken in the large-deviation sense, and $\dd[N]{y^0} \dd[N]{\theta}$ are transformed into integration measures on densities. But if we neglect the $e^{\mathcal O(1)}$ correction, then this {\em does not} satisfy the normalisation \eqref{rhopnorm}, because of the missing space in $\Sigma$. Hence this break conservation of probability (or unitarity), which is a crucial fundamental concept in quantum mechanics. Note that the missing space in $\Sigma$ is {\em macroscopic}, as it encodes the macroscopic effects of interactions on the positions of particles at macroscopic times in a finite-density gas. Possibly, at least part of the $e^{\mathcal O(1)}$ corrections, coming from linearisation of the phase, could take care of the missing space, but currently we do not know how to show this. Further, transforming the $+\mathcal O(1/\ell)$ in the second argument of $\b {\mathrm W}$ in \eqref{equ:LL_deriv_2_deriv_3} into $e^{\mathcal O(1)}$, that is $\b{\mathrm{W}}_{N}(\bs{q},\bs{\rapidity}+\tfrac{1}{2\ell}\bs{q}) \approx \b{\mathrm{W}}_{N}(\bs{q},\bs{\rapidity})e^{\mathcal O(1)}$, assumes that $\b{\mathrm{W}}_{N}(\bs{q},\bs{\rapidity})$ had variations of order $1/\ell$ on each rapidity argument in a region $\rapidity_i + \order{1/\ell}$. However, it seems reasonable to us that $\b{\mathrm{W}}_{N}(\bs{q},\bs{\rapidity})$ is a rather rough function (on this scale) in the second component due to the antisymmetry of the amplitude; indeed, the antisymmetry of $A(\bs\theta)$ translates into a complicated condition for the Wigner distribution, which, for $\mathcal O(N) = \mathcal O(\ell)$ arguments, may produce this roughness.

Note that the Bethe states \eqref{equ:pre_int_LL_Nparticle} are only an orthonormal basis if appropriately antisymmetrized by the sum over permutations. Also, they are only orthonormal if $\phi(\rapidity)$ is precisely the Lieb-Liniger phase shift. Hence, a derivation of GHD should make use of the permutation sum and the precise form of the phase-shift in some way, otherwise violations of unitarity, as appears to be the case from \eqref{equ:LL_deriv_2_final_hatx}, are expected.

Nevertheless, even though the derivations so far are flawed, they clearly demonstrate the main principle how GHD emerges in quantum integrable models like the Lieb-Liniger model: it emerges via a stationary phase approximation, or somewhat equivalently a phase linearisation, giving rise to a classical particle model. How to properly justify its emergence in terms of expectation values requires a proper treatment of the fermionic nature of the quasi-particles, i.e.\ the sum over the permutations. 

\begin{remark} In \eqref{equ:LL_deriv_2_final_hatx}, the change of coordinates may have been implemented by a Jacobian, the determinant of
\begin{align}
    \partial_{x_j}\b y_i(\bs{x},\bs{\rapidity}) &= \delta_{ij} + \tfrac{1}{\ell} \sum_{j\neq i}\delta(x_i-x_j) \varphi(\rapidity_i-\rapidity_j).\label{equ:LL_deriv_2_final_jacobian}
\end{align}
However, this is of course extremely singular due to the $\delta$ functions. \change{However, if one regularizes the $\delta$ function by a smooth function, the same determinant is the Gaudin determinant of the corresponding classical systems~\cite{PhysRevLett.132.251602,doyon2023generalisedtbartdeformationsclassicalfree}. In fact, it can be computed exactly using the matrix tree theorem~\cite{doyon2023generalisedtbartdeformationsclassicalfree}. We note that this determinant also plays a crucial role in the thermodynamics of the classical models, where it gives rise to the interaction term. In fact, it appears also due to a coordinate change $y\leftrightarrow x$, however, it appears in the other direction.}
\end{remark}

\section{Heisenberg picture}\label{sec:heisenberg}

In the previous sections, we saw how the Bethe ansatz structure of the Lieb-Liniger eigenfunctions gives rise, in natural ways, to a mechanical system of particles in interaction, which itself implies the GHD equation. This system may be interpreted as describing the positions of Bethe ansatz wave packets as they evolve under the Schr\"odinger equation. One problem, which we have not resolved, is that these wave packets do not take into account the fermionic statistics. Hence, there is a difficulty to relate these somewhat abstract wave packets positions to actual averages of local observables, and in particular, the average of the operator measuring the spectral phase-space density -- the principal object in GHD.

In this section, we show that, in fact, {\em the quantum version of this mechanical system of particles follows from the Bethe ansatz structure}, without the need for large-scale approximation into wave packets, or Wigner functions. That is, the quantisation of the Bethe system \eqref{equ:LL_SP_QP_def} arises, without approximations, as an operator relation. In this operator relation, $x_i$  is replaced by the quantum-mechanical position operator $\hat x_i$, and $\theta_i,\,y_i$ by operators $\hat \theta_i,\,\hat y_i$ that represent the mapping to free-motion of the quantum mechanical momentum and position operators of the interacting model. As a consequence, the spectral phase-space density operator satisfies the operator version of the GHD equation.

The question of the emergence of the GHD equation is therefore recast into that of taking the average of the operatorial GHD equation -- again we do not solve this problem, but provide a general discussion.

\subsection{Quantised soliton gas}

Consider the (un-normalisable, un-symmetrized) wave functions in the Hilbert space $L^2(\R^N)$
\beq\label{equ:unormalized_chi}
	\chi(\bs x|\bs\theta) = (2\pi)^{-N/2}s(\bs x)e^{\ri \Phi(\bs x,\bs \theta)}
\eeq
where $\Phi(\bs x,\bs \theta)$ is defined in \eqref{equ:pre_int_LL_phase}.
Our main assumption is that $\{\chi(\cdot|\bs\theta):\bs\theta\in \R^N\}$ forms a basis of $L^2(\R^N)$ in the spectral sense. This basis is not expected to be orthogonal, but this is not important for our discussion. We use the notation $|\bs\theta\rangle$ for the corresponding vector, with $\langle \bs x|\bs\theta\rangle = \chi(\bs x|\bs\theta)$. Clearly, the Lieb-Liniger eigenfunctions \eqref{equ:pre_int_LL_Nparticle}, element of the fully symmetric subspace  $\mathcal S L^2(\R^N)\subset L^2(\R^N)$, are written as
\begin{equation}
    \psi(\bs x|\bs\theta)
    =
    \frc{1}{\sqrt{N!}}\sum_{\sigma\in\mathcal S_N}(-1)^{|\sigma|}\chi(\bs x_{\sigma},\bs\theta).
\end{equation}

We define the operators $\h\theta_i$ for $i=1,2,\ldots,N$ as follows:
\beq\label{deftheta}
	\h\theta_i |\bs\theta\rangle = \theta_i|\bs\theta\rangle\ \forall\ \bs\theta\in \R^N.
\eeq
Because $\chi(\cdot|\bs\theta)$'s do not form an orthonormal basis, the operators $\hat\theta_i$ are not Hermitian. Nevertheless they are well-defined operators on $L^2(\R^N)$.

The Lieb-Liniger model is a quantum model for Bosons, hence it is defined on the fully symmetric Hilbert space $L^2\ind{sym}(\R^N)$. Clearly, the Hamiltonian \eqref{equ:LL_Hamiltonian} preserves this subspace of $L^2(\R^N)$, and it is its action on this space that defines the time evolution of the model. Although each $\h\theta_i$ does not preserve this subspace, the Hamiltonian may still be written in terms of these operators:
\beq\label{Htheta}
    \hat H|_{L^2\ind{sym}(\R^N)} = \sum_i \frc{\hat \theta_i^2}2\quad:\quad L^2\ind{sym}(\R^N) \to L^2\ind{sym}(\R^N).
\eeq
Indeed, the sum over all particles preserves $L^2\ind{sym}(\R^N)$. This is shown as follows. Let $\psi\in L^2\ind{sym}(\R^N)$. Then we may expand it as
\beq\label{psigeneric}
    \psi(\bs x) = 
    \int \dd^N\theta \,A(\bs\theta)\chi (\bs x|\bs\theta)
    =
    \frc1{N!}\sum_{\sigma\in\mathcal S_N}
    \int \dd^N\theta \,A(\bs\theta)\chi (\bs x_\sigma|\bs\theta)
    =
    \frc1{N!}\sum_{\sigma\in\mathcal S_N}
    \int \dd^N\theta \,A(\bs\theta_\sigma)\chi (\bs x|\bs\theta)
\eeq
where in the second equality we have used the fact that the function is assumed to be fully symmetric, and in the last we have used $\chi (\bs x_\sigma|\bs\theta) = \chi (\bs x|\bs\theta_{\sigma^{-1}})$ and then a change of integration variable. Hence
\beqa
    \sum_i\frc{\h\theta_i^2}2
    \psi (\bs x)
    &=&
    \frc1{N!}\sum_{\sigma\in\mathcal S_N}
    \int \dd^N\theta\,
    \sum_i\frc{\theta_{i}^2}2
    A(\bs\theta_\sigma)\chi (\bs x|\bs\theta)
    \n &=&
    \frc1{N!}
    \sum_{\sigma\in\mathcal S_N}
    \int \dd^N\theta \,\sum_i\frc{\theta_{\sigma(i)}^2}2
    A(\bs\theta_\sigma)\chi (\bs x|\bs\theta)
\eeqa
where in the second equality we have changed variable $i\to \sigma(i)$. We see that the right-hand side has the form \eqref{psigeneric} with expansion coefficient $\sum_i \theta_i^2/2\,A(\bs \theta)$, and thus it is fully symmetric.

Crucially, expression \eqref{Htheta} then allows us to {\em extend the Hamiltonian to an operator acting on $L^2(\R^N)$} in the following way:
\beq
    \hat H' = \sum_i \frc{\hat \theta_i^2}2  \quad :\quad L^2(\R^N)\to L^2(\R^N)
\eeq
with
\beq
    \hat H = \hat H'\quad \mbox{on}\quad \mathcal L^2\ind{sym}(\R^N).
\eeq
Note that this is in general {\em different from \eqref{equ:LL_Hamiltonian} seen as an operator on $L^2(\R^N)$}: $\hat H' \neq \hat H$ on $L^2(\R^N)$. Further, $\h H'$ is not expected to be Hermitian on $L^2(\R^N)$. But this is not important, because, as recalled, the Lieb-Liniger model is defined as a quantum mechanical model on $L^2\ind{sym}(\R^N)$, and on this space the operators agree.

We note that the spectral phase-space density operator, defined on $L^2\ind{sym}(\R^N)$ in \eqref{equ:LL_deriv_cq_def}, can be written to $L^2(\R^N)$ simply by using the operators $\h\theta_i$:
\beq\label{densityoperator}
    \h\rho(x,\theta)=\sum_i \delta(x-\h x_i)\delta(\theta-\h\theta_i).
\eeq

Now consider the functions
\beq
	y_i(\bs x,\bs\theta) = x_i + \frc12\sum_{j\neq i}\sgn(x_i - x_j)\varphi(\theta_i- \theta_j),\quad \varphi(\theta) = \frc{\p\phi(\theta)}{\p\theta}
\eeq
which encode the change of variable \eqref{equ:LL_SP_QP_def}. For fixed $\bs \theta$ and with $\sgn$ chosen as a smooth regularisation of the sign function, these are smooth diffeomorphisms $\bs x \mapsto y_i(\bs x,\bs\theta)$. We also define the smooth functions
\beq\label{yxtheta}
	y_{\bs x,\bs\theta}(x,\theta) =  x + \frc12\sum_{j}\sgn(x-x_{j})\varphi(\theta-\theta_{j})
\eeq
and clearly
\beq\label{yirelation}
	y_i(\bs x,\bs\theta) = y_{\bs x,\bs\theta}(x_i,\theta_i).
\eeq
We define the operators $\h y_i$ for $i=1,2,\ldots,N$ as follows:
\beq\label{defy}
	\h y_i = \nor y_i(\h{\bs x},\h{\bs\theta})\nor
	=\h x_i + \frc12\sum_{j\neq i}\sgn(\h x_i - \h x_j)\varphi(\h \theta_i - \h\theta_j).
\eeq
Here we have introduced the {\em classical ordering}, defined by putting the $\h x_i$'s on the left of the $\h\theta_i$'s,
\begin{equation}\label{nor}
    \nor \hat \theta_i \hat x_{j}\nor
    =\hat x_j\hat \theta_i,\quad \mbox{etc.}
\end{equation}
It is simple to verify that
\beq
	\h y_i\chi(\bs x,\bs\theta) = -\ri \frc{\p\chi(\bs x,\bs\theta)}{\p\theta_i}
\eeq
and hence
\beq
	[\h y_i,\h\theta_j] = \ri \,\delta_{ij}.
\eeq
Because of this commutation relation, we have
\beq\label{evoly}
    e^{\ri \hat H't}\h y_i e^{-\ri \hat H't} = \h y_i + \ri \h \theta_i t.
\eeq

Hence, combining \eqref{defy} and \eqref{evoly}, we find the {\em quantum version of the Bethe system of particles} \eqref{equ:LL_SP_QP_def},
\beq\label{equ:LL_SP_QP_def_op}
    \h y_i + \h\theta_i t
    = \h x_i(t) + \frc12
    \sum_{j\neq i}\sgn(\h x_i(t)-\h x_j(t))\varphi(\h\theta_{ij}).
\eeq
This is an exact operatorial relation, and in particular it does not require any large-scale approximation, in contrast to \eqref{equ:LL_SP_QP_def}.

\subsection{Height-field method for showing the emergence of the GHD equation}\label{ssectheighfieldmethod}

Taking a time derivative of \eqref{equ:LL_SP_QP_def} it is a simple matter to arrive, by assuming a certain large-scale form of $\dot x_i(t)$, at the effective velocity, and hence at the GHD equation; see the derivation around Eq.~\eqref{equ:veff_deriv}, and \cite{PhysRevLett.132.251602}. However, for the operator relation \eqref{equ:LL_SP_QP_def_op}, these steps require modifications because derivatives of functions of operators are more complicated. The  calculation is much less well justified. An alternative derivation of the GHD equation from \eqref{equ:LL_SP_QP_def} is presented in \cite{PhysRevLett.132.251602}, which could also be adapted to the quantum case, but operator-ordering problems remain.

A more powerful method to obtain the GHD equation is that based on height fields. As shown in \cite{hübner2024newquadraturegeneralizedhydrodynamics,hübner2024existenceuniquenesssolutionsgeneralized}, an equivalent formulation of the GHD equation is a self-consistent, fixed-point problem for the quasi-particle height field,
\beq\label{defNGHD}
    N(t,x,\theta) = \frc12\int\dd x'\,\sgn(x-x')\rho(t,x',\theta),\quad
    \rho(t,x,\theta) = \p_x N(t,x,\theta).
\eeq
The fixed-point problem is expressed as follows: there exists a function $M(x,\theta)$ such that
\beq\label{fixedpoint}
    N(t,x,\theta)
    =
    M\Big(x-\theta t+\int \dd\theta'
    N(t,x,\theta')\varphi(\theta-\theta'),\,\theta\Big).
\eeq
The function $M(x,\theta)$ is fully determined by the initial condition, simply by taking the case $t=0$ of the fixed-point problem. Once it is determined, the fixed-point problem then determines $N(t,x,\theta)$ for all times $t$, and by differentiation, the spectral phase-space density.

It turns out that the fixed-point problem \eqref{fixedpoint} has both {\em an exact form in terms of the semi-classical Bethe system \eqref{equ:LL_SP_QP_def}}, and, again, {\em an exact operatorial form in the Lieb-Liniger model}, both obtained without large-scale approximation. These are more useful forms of the GHD equation in order to prove rigorously that it emerges at large scales.

We now show how both fixed-point problems, for the semi-classical Bethe system, and as an operatorial equation in the Lieb-Liniger model, are obtained. We will not show rigorously how the GHD equation is then obtained from this, but provide a general discussion.

Let $x_i(t)$ satisfy the semiclassical Bethe system \eqref{equ:LL_SP_QP_def}, and define the empirical height function
\beq\label{defNcl}
	N_{\bs x,\bs\theta}(t,x,\theta) = \frc12\sum_j \sgn(x-x_j(t))\delta(\theta-\theta_j).
\eeq
Here and below, we implicitly see each $x_j(t) = x_j[\bs x,\bs\theta](t)$ as a function of the (time-0) coordinates $\bs x,\bs\theta$ via the solution to \eqref{equ:LL_SP_QP_def}. With the empirical density
\beq
    \rho_{\bs x,\bs\theta}(t,x,\rapidity) = \sum_i \delta(x-x_i(t))\delta(\rapidity-\rapidity_i)
\eeq
it is clear that relation \eqref{defNGHD} holds for these quantities. With \eqref{yxtheta}, we have
\beq\label{yxtN}
	y_{\bs x(t),\bs \theta}(x,\theta) = x + \int \dd\theta'\,N_{\bs x,\bs\theta}(t,x,\theta') \varphi(\theta-\theta')
\eeq
and by the dynamics \eqref{equ:LL_SP_QP_def} along with \eqref{yirelation},
\beq
    y_i(\bs x,\bs\theta) + \theta_i t = y_{\bs x(t),\bs \theta}(x_i(t),\theta_i) = x_i(t) + \int \dd\theta\,
    N_{\bs x,\bs\theta}(t,x_i(t),\theta)\,\varphi(\theta_i-\theta).
\eeq
As $y_{\bs x,\bs \theta}(x,\theta)$ is monotonically increasing in $x$, we have
\beq\label{sgnrel}
	\sgn(y_{{\bs x}(t),{\bs \theta}}(x,\theta_i) - y_i(\bs x,\bs\theta)-\theta_i t)
	=
	\sgn(x-x_i(t)).
\eeq
Define the empirical height function for the free coordinates,
\beq
	M_{\bs x,\bs\theta}(y,\theta) = \frc12\sum_j \sgn(y-y_j(\bs x,\bs\theta))\delta(\theta-\theta_j).
\eeq
Then by \eqref{sgnrel} and \eqref{defNcl},
\beq\label{MN}
    M_{\bs x,\bs\theta}(y_{\bs x(t),\bs\theta}(x,\theta)-\theta t,\theta) = N_{\bs x,\bs\theta}(t,x,\theta)
\eeq
and using the relation \eqref{yxtN} we obtain
\beq\label{fixedpointempirical}
	N_{\bs x,\bs\theta}(t,x,\theta) = M_{\bs x,\bs\theta}\Big(x -\theta t + \int \dd\theta'\, N_{\bs x,\bs\theta}(t,x,\theta') \varphi(\theta-\theta'),\,\theta\Big),
\eeq
which is the equivalent of \eqref{fixedpoint} where the GHD height function is replaced by the empirical height function, defined within the semiclassical Bethe system of classical particles. Clearly, if both empirical height functions $N_{\bs x,\bs \theta}$ (for the real positions) and $M_{\bs x,\bs \theta}$ (for the free-space positions) have a limit  in the large-scale regime where $x_i = \ell \b x_i,\,t = \ell \b t,\,\ell\to\infty,\,N\to\infty$, in an appropriate topology, then we directly obtain the fixed-point formulation of the GHD equation.

We now claim that the same relation holds in the Lieb-Liniger model for the corresponding quantum operator height functions, under an appropriate normal ordering. That is, define
\beq\label{defN}
	\h N(t,x,\theta) = \frc12\sum_j \sgn(x-\h x_j(t))\delta(\theta-\h\theta_j)
\eeq
and
\beq
	\h M(y,\theta) = \frc12\sum_j \sgn(y-\h y_j)\delta(\theta-\h\theta_j)
\eeq
where the $\h y_j$ operators are defined in \eqref{defy} and satisfy \eqref{equ:LL_SP_QP_def_op}. Then
\beq\label{mainrelation}
	\h N(t,x,\theta) = \nor \h M\Big(x -\theta t + \int \dd\theta'\, \h N(t,x,\theta') \varphi(\theta-\theta'),\,\theta\Big)\nor_t
\eeq
where the $t$-classical ordering generalises \eqref{nor} to the time-evolved coordinates,
\beq
    \nor \h\theta_i\h x_j(t)\nor_t
    =
    \h x_j(t)\,\h\theta_i,\quad \mbox{etc.}
\eeq
We now show \eqref{mainrelation} directly (this can be adapted to an alternative proof of \eqref{fixedpointempirical}).

In order to do so, we must write the operator $\h M\Big(x -\theta t + \int \dd\theta'\, \h N(t,x,\theta') \varphi(\theta-\theta'),\,\theta\Big)$ in the ``alphabet'' of $\h x_j(t)$'s and $\h\theta_j$'s, and apply the $t$-normal ordering. We note that there is a unique way of writing an operator in this alphabet in the $t$-normal ordered form. Given an operator $\h A$ on $L^2(\R^N)$, this takes the form
\beq
    \nor\h A\nor_t = \int \dd^N x\dd^N\theta\,
    \langle \bs x|e^{-\ri \h H' t}
    \nor\h A\nor_t|\bs \theta\rangle
    \delta^N(\bs x-\h{\bs x}(t))
    \delta^N(\bs \theta-\h{\bs \theta}).
\eeq
as both sides have the same value under $\langle \bs x'|e^{-\ri \h H' t}\cdot|\bs\theta'\rangle$ for all $\bs x'$ and $\bs\theta'$, and $\langle \bs x'|e^{-\ri \h H' t}$ and $|\bs\theta'\rangle$ both form a basis. Using \eqref{equ:LL_SP_QP_def_op} we have
\beq
    x - \h y_i - \h\theta_i t + \int \dd\theta'\, \h N(t,x,\theta') \varphi(\theta-\theta')
    =
    x-\h x_i(t)
    +\frc12\sum_{j\neq i}\big(
    \sgn(x-\h x_j(t))
    -
    \sgn(\h x_i(t)-\h x_j(t))\big)
    \delta(\theta-\h\theta_j)
\eeq
which implies
\beqa
    \lefteqn{\langle\bs x| e^{-\ri \h H't}\nor \h M\Big(x -\theta t + \int \dd\theta'\, \h N(t,x,\theta') \varphi(\theta-\theta'),\,\theta\Big) \nor_t
    |\bs\theta\rangle}
    && \\
    &=&
    \frc12\sum_{i}
    \sgn\Big(
    x-x_i(t)
    +\frc12\sum_{j\neq i}\big(
    \sgn(x-x_j(t))
    -
    \sgn(x_i(t)-x_j(t))\big)
    \delta(\theta-\theta_j)
    \Big)\delta(\theta-\theta_i)
    \no
\eeqa
By monotonicity,
\beq
    \sgn(x-x_j(t))
    -
    \sgn(x_i(t)-x_j(t))\quad
    \geq 0 \mbox{ if } x\geq x_i(t),\quad
    \leq 0 \mbox{ if } x\leq x_i(t).
\eeq
and therefore
\beqa
    \langle\bs x| e^{-\ri \h H't}\nor \h M\Big(x -\theta t + \int \dd\theta'\, \h N(t,x,\theta') \varphi(\theta-\theta'),\,\theta\Big) \nor_t
    |\bs\theta\rangle
    &=&
    \frc12\sum_{i}
    \sgn(
    x-x_i(t))
    \delta(\theta-\theta_i)\n
    &=&
    \langle\bs x| e^{-\ri \h H't}
    \h N(t,x,\theta)
    |\bs\theta\rangle
\eeqa
which implies \eqref{mainrelation} as $\h N(t,x,\theta)$ is by definition $t$-classical ordered.

\subsection{Discussion: commutativity of fluid-cell means of charge densities}

The operatorial formulation \eqref{mainrelation} of the GHD equation does not rely on any properties of the state. \change{However, it is not yet sufficient in order to obtain the GHD equation. We need to evaluate \eqref{mainrelation} within a long-wavelength state, taking $x,\,t = \ell \b x,\,\ell\b t$, and show that it becomes, in the macroscopic limit \eqref{equ:macrolimit}, the ``same'' equation but for the corresponding classical quantities. This will happen if (1) the state satisfies a strong enough clustering property, (2) Eq.~\eqref{mainrelation} can be recast in term of fluid-cell means (i.e.~mesoscopic spatial means) of $\hat\rho(x,\theta)$ (Eq.~\eqref{densityoperator}), and (3) such fluid-cell means commute. Then, we would obtain the universal GHD equation for the factorised average $\langle \h\rho(x,\theta)\rangle$.}


\change{Steps (1) and (2) are difficult steps, which we will not develop here. However, we can show point (3), which also reveals an interesting structure related to semiclassical analysis for interacting quantum models.}

The spectral phase-space density operator {\em at large scales} is its macroscopic version
\begin{equation}
    \h\rho[f] = \int \dd x\dd\theta\,\h\rho(x,\theta) f(x,\theta),
\end{equation}
with slowly-varying function $f(x,\theta) = \b f(x/\ell,\theta)$, where $\b f$ is an $\ell$-independent rapidly decreasing function. This is a way of implementing the fluid-cell mean of $\h\rho(x,\theta)$. It is simple to show the phenomenon of ``classicalisation'': {\em the large-scale operator $\h\rho[f]$ becomes essentially classical as $\ell\to\infty$}, in that $\h\rho[f]$ for different $f$'s, and also their Hermitian conjugates, commute strongly enough with each other at large $\ell$.

The operator $\h\rho[f]$ is $\ell$-extensive: it is the integral over a local operator on an interval of length essentially of order $\ell$. Thus its average, for finite densities, is $\mathcal O(\ell)$. Commutators of extensive operators are also extensive, by the fact that local operators at distances far enough from each other commute. But here we have more: we will show
\begin{equation}\label{rhorhodag}
    [\,\h\rho[f],\h\rho[g]\,],\quad  [\h\rho[f],\h\rho^\dag[g]]\quad\mbox{are $\ell$-extensive observables with densities that are $\mathcal O(\ell^{-1})$}.
\end{equation}
Therefore, this property subsists for all commutators.

Eq.~\eqref{rhorhodag} follows from two basic ingredients. First, we use the locality of the (not necessarily Hermitian) conserved densities $\h{ q}_a = \int \dd \theta\,\theta^a\h\rho(x,\theta)$; this was argued for in Appendix \ref{app:densities_of_cqs}. Second, we use the fact that local conserved densities for conserved quantities that are in involution $[\h Q_a,\h Q_b]=0\;\forall\;a,b$, as is the case for the LL model's $\h Q_a = \int \dd x\,\h{\mathcal q}_a(x)$, satisfy a very special commutation relation, as established in \cite{Doyon2017note} (see Eqs.~(92), (93) there):
\begin{equation}\label{maincommutator}
\begin{aligned}{}
    [\h{ q}_a(x),\h{ q}_b(x')] &=
    \ri \p_{x'} \h{ j}_{ab}(x')\delta(x-x') + \h u(x')\delta'(x-x')
    + \ldots\\
    [\h{ q}_a^\dag(x),\h{ q}_b(x')] &=
    \ri \p_{x'} \h{ j}_{ab}(x')\delta(x-x') + \h v(x')\delta'(x-x')
    + \ldots
    \end{aligned}
\end{equation}
where $\h{ j}_{ab}(y)$ is the (generalised) current satisfying
\begin{equation}
    \ri [\h Q_a,\h{ q}_b(x)] + \p_x \h{ j}_{ab}(x) = 0.
\end{equation}
Note in particular that $\h{ j}_{0b} =0$, $\h{ j}_{1b} =  \h{ q}_b$ and $\h{ j}_{2b} = 2 \h{ j}_b$ where $\h{ j}_b$ is the usual current. In \eqref{maincommutator}, $\h u(x'),\,\h v(x')$ are some local operators (see \cite[Eqs 92, 93, 95]{Doyon2017note}) and ``$\ldots$'' represents similar terms with higher derivatives of the delta function, $\delta^{(n)}(x-y)$, $n>1$. Here, in fact, we have extended the arguments of \cite{Doyon2017note} to include the commutator with Hermitian conjugates, using the fact that $\h Q_a = \int \dd x\,\h{ q}_a(x) = \int \dd x\,\h{ q}_a^\dag(x)$ is Hermitian. With this, we write
\beq
    f(x,\theta) = \sum_a f_a(x)\theta^a,\quad
    g(x,\theta) = \sum_a g_a(x)\theta^a
\eeq
and obtain
\begin{align}
    [\,\h \rho[f],\h\rho[g]\,] &=
    \sum_{ab} \int \dd  x\dd {x}'\,  f_a( x)g_{b}( x')[\h{ q}_a(x),\h{ q}_b(x')]\n
    &= \sum_{ab} \Big[-\ri \int \dd x\,\p_x (f_a(x)g_b(x)) \h{ j}_{ab}(x)
    -
    \int \dd x\,\p_x f_a(x) g_b(x) \h u(x) + \ldots\Big]
\end{align}
where ``$\ldots$" involve higher-order derivatives. Using $f_a(x) = \b f_a(x/L)$, $g_a(x) = \b g_a(x/L)$ with $\b f_a$ and $\b g_a$ being fixed rapidly decreasing functions, we obtain the first part of Eq.~\eqref{rhorhodag}. The second \eqref{rhorhodag} is proven in a similar way.

We can go further and evaluate exactly the density of the resulting extensive operator. We use the fact that \cite[Eq 95]{Doyon2017note}
\begin{equation}
    \h u_{ab}(x) = -\ri (
    \h{ j}_{ab}(x) +
    \h{ j}_{ba}(x)
    ) + \p_x \h w(x)
\end{equation}
for some local operator $\h w(x)$, and obtain
\begin{equation}
    \label{commutatorgen}
    [\,\h \rho[f],\h\rho[g]\,]
    =
    \ri \sum_{ab}\Big[
    \p_x f_a(x)g_b(x)\hat{ j}_{ba}(x)
    - f_a(x)\p_x g_b(x)
    \hat{ j}_{ab}(x)
    \Big]
    + \mathcal O(\ell^{-1})
\end{equation}
where $\mathcal O(\ell^{-1})$ means that within a state, the average scales in this way. We note that at the Tonks-Girardeau point, we have a free-fermion structure and $\h{\mathcal j}_{ab}(x) = \int \dd \theta\,\theta^{b}v_a(\theta)\h\rho(x,\theta)$ where the $a$-velocity is $v_a(\theta) = \p_\theta \theta^a = a\theta^{a-1}$ \cite{Doyon2017note}. This then gives
\begin{equation}
    [\,\h \rho[f],\h\rho[g]\,]
    =
    \ri \h\rho[\{f,g\}] + O(\ell^{-1})\qquad\mbox{(Tonks-Girardeau point)}
\end{equation}
where $\{\cdot,\cdot\}$ is the classical Poisson bracket.
But Eq.~\eqref{commutatorgen} holds for arbitrary quantum models, integrable or not, with an arbitrary set of conserved quantities $\h Q_a$ in involution. We believe it can form the basis for a generalisation of semiclassical analysis \cite{Martinezbook,Zworskibook} to interacting models. We will analyse this in a future work.

\section{Conclusion}
We have made important steps towards an ab initio derivation of the equations of GHD in interacting quantum models, focusing on the Lieb-Liniger model. Crucially, we have identified how the evolution of the quantum mechanical wave function is connected to GHD: the connection is made via an intermediate classical particle model, describing the evolution of wave packets on the quantum side and the evolution of the quasi-particles on the GHD side. Interestingly, this shows that the density of wave packets also satisfies the GHD equation even though it differs from the physical density. We have also established an explicit formula (Appendix \ref{app:densities_of_cqs}) for how the local densities of the conserved quantities act on Bethe wave functions. Our main result is that this very natural formula gives, indeed, {\em local} densities. We have explored potential routes of how to derive the GHD equation for the physical density as well, both from the Schrödinger and from the Heisenberg picture. While we believe all of these routes can give rise to proper derivations of GHD, in their current form gaps remain. Compared to classical systems, mathematical frameworks are much less developed for quantum systems. Since, additionally, we lack the numerical tools to verify the approximations made, we have to rely on physical intuition.

We thus would like to point out the great need to develop appropriate mathematical frameworks, for instance: a) analogs of large deviation theory that can describe oscillating integrals and can take particle statistics into account, or b) methods to understand when and how normal-ordering can be dropped in the Heisenberg picture.

With the advance of quantum computers in mind, it would also be interesting to verify (or falsify) the validity of approximations like \eqref{equ:LL_SP_SP_WF}, or \eqref{Psimodulatedinteractingtime} with \eqref{modulationevolve}, on a quantum computer.

Once these derivations are better understood, they could serve as a starting point for gaining deeper understanding into the large-scale behaviour of quantum integrable systems. Specifically, it would be interesting to systematically understand higher-order corrections, such as the diffusive corrections. Recently it was proposed, and checked numerically, in \cite{PhysRevLett.134.187101} that diffusive corrections are obtained from the ballistic transport of initial fluctuations, as per \cite{PhysRevLett.131.027101,10.21468/SciPostPhys.15.4.136}, under a universal regularisation scheme. This followed related works \cite{PhysRevB.109.024417,PhysRevLett.128.160601,PhysRevE.111.024141,doi:10.1073/pnas.2403327121}. We note that the absence of emergent hydrodynamic noise in integrable systems was explained and argued to hold at all orders in \cite{doyon2025hydrodynamicnoisedimensionprojected}, and the fact that initial fluctuations don't affect coarse-grained quantities was analysed and numerically confirmed at the diffusive order of the hard-rod model in \cite{hubner2025hydrodynamicsaveraginghard}. There is the proposal~\cite{urilyon2025simulatinggeneralisedfluidsinteracting} that such diffusive (and potentially also higher order) corrections can be correctly taken into account by initializing the corresponding semi-classical Bethe model with states that have the same (coarse-grained) correlations as quantum states. Such higher order corrections are purely of classical nature. It is clear that quantum correlations must also eventually affect the dynamics. However, since in the thermodynamic limit quantum fluctuations are suppressed, still relatively little is known about how this happens, and how to adjust the hydrodynamic equations to take them into account \cite{Alba_2021}. Only in the low temperature regime, where classical fluctuations are suppressed as well, there is a theory called quantum GHD~\cite{PhysRevLett.124.140603}, which asserts that quantum correlations are transported ballistically. In Quantum GHD one quantizes the hydrodynamic excitations, i.e. sound modes, similar to Luttinger liquid theory. Extending the derivations done in this paper in this regime and identifying the quantum corrections will therefore help to gain crucial insights into the validity of this phenomenological quantum theory.

\textbf{Acknowledgements}
The authors are grateful to Alvise Bastianello, Thibault Bonnemain, Olalla Castro Alvaredo and Jacopo De Nardis for discussions. The work of BD was supported by the Engineering and Physical Sciences Research Council (EPSRC) under grants no EP/W010194/1 and EP/Z534304/1 (ERC advanced grant scheme). BD would like to thank the Isaac Newton Institute for Mathematical Sciences, Cambridge, for support and hospitality during the programmes ``Dispersive hydrodynamics: mathematics, simulation and experiments, with applications in nonlinear waves" and ``Building a bridge between non-equilibrium statistical physics and biology" where work on this paper was undertaken (EPSRC grant no EP/R014604/1). FH acknowledges funding from the faculty of Natural, Mathematical \& Engineering Sciences at King’s College London.

\appendix

\section{Integrable models with strings: attractive Lieb-Liniger}\label{app:attractive}
In this appendix, we would like to explain how the derivation in section \ref{sec:derivation_wavepackets} can be extended to models with bound states (called strings). We will use the attractive Lieb-Liniger model, i.e. $c<0$ as an example.

We will start by briefly summarizing why and how string states appear~\cite{10.1063/1.1704156,jscaux}. The idea is that for $c<0$ Bethe states like \eqref{equ:pre_int_LL_Nparticle} are still eigenstates, however, it is possible that the rapidities $\rapidity_i$ are complex numbers. To see this, let us restrict \eqref{equ:pre_int_LL_Nparticle} to the sector $x_1<x_2 <\ldots <x_N$ and write \eqref{equ:pre_int_LL_Nparticle} as
\begin{align}
		\psi(\bs{x}|\bs{\rapidity}) \sim \sum_{\sigma\in \mathcal{S}_N} (-1)^{|\sigma|} e^{i\sum_i\rapidity_{\sigma_i}x_i} \prod_{i<j} B(\rapidity_{\sigma_j}-\rapidity_{\sigma_i}),\label{equ:LL_attractive_Bethe}
\end{align}
where $B(\rapidity) = e^{\tfrac{i}{2}\phi(\rapidity)}\sqrt{c^2+\rapidity^2} = c+i\rapidity$. To ensure that there is no blowup as $x_1\to -\infty$, we either require $\Im \rapidity_{\sigma_1} \leq 0$ or $\prod_{i<j} B(\rapidity_{\sigma_j}-\rapidity_{\sigma_i}) = 0$. This implies that if any $\Im \rapidity_i > 0$, there has to be another particle $j$ such that $B(\rapidity_j-\rapidity_i) = 0$, implying $\rapidity_j = \rapidity_i + ic$. Similarly, by sending $x_N \to \infty$ we find that if $\Im \rapidity_i < 0$, then there must be another $\rapidity_j = \rapidity_i - ic$. This is only possible for $c<0$, where additional solutions of the form~\cite{10.1063/1.1704156,jscaux}
\begin{align}
	\rapidity_{a,k} &= \rapidity_a + ik\abs{c} & & k=-(l-1)/2, -(l-1)/2 +1 ,\ldots, (l-1)/2,\label{equ:LL_attractive_string_rapidities}
\end{align}
called strings of length $l = 1, 2, 3, \ldots$, are present.  Here, $\rapidity_a\in \mathbb{R}$ is the center of the string. Strings of length $l \geq 2$ are bound states of particles and are interpreted as new types of quasi-particles with asymptotic momentum $P_a = \sum_{k = -(l-1)/2}^{(l-1)/2} \rapidity_{a,k} = l \rapidity_a$ and energy $E_l = \sum_{k = -(l-1)/2}^{(l-1)/2} \rapidity_{a,k}^2 = l \rapidity_a^2 - \abs{c}^2 \sum_{k = -(l-1)/2}^{(l-1)/2} k^2$~\cite{10.1063/1.1704156,jscaux}. These can also be observed experimentally~\cite{horvath2025observingbethestringsattractive}). 

\begin{remark}
While the structure of strings is simple on the infinite line, it becomes incredibly more complicated on a finite system of size $\ell$ (with say periodic boundary conditions). On a finite system, instead of simply decaying as $x \to \pm \infty$, they need to be periodic. This periodicity quantizes the momenta in complicated ways, and explicit solutions are not generally known on a finite system. In the thermodynamic limit $\ell \to \infty$, however, it is generally expected that strings will be well described by the infinite line structure \eqref{equ:LL_attractive_string_rapidities}. This is the string hypothesis~\cite[Chap 6]{arutyunov2020elements}. This can easily be understood as follows. The typical size of a string is of order $\sim \abs{c}$, meaning that the wave function corresponding to an infinite line string \eqref{equ:LL_attractive_string_rapidities} will be an eigenstate of \eqref{equ:LL_Hamiltonian}, up to an exponential small error $\sim e^{-\ell/\abs{c}}$.
\end{remark} 

Note that in the standard derivation of hydrodynamics via thermodynamic properties, it is crucial to first study finite system sizes $\ell$ and then send $N\sim \ell \to \infty$ simultaneously. Hence, the string hypothesis is a crucial assumption in this derivation. The ab initio derivation proposed in this paper immediately works on the infinite system, hence no string hypothesis is required.

In the following, let us label particles as $i \to (l,a,k)$, where $l$ is the string length, $a=1, \ldots, N_l$ is its label and $k=-(l-1)/2, \ldots, (l-1)/2$ is the $k$ particle in $a$. We will denote the individual rapidities as $\rapidity_{l,a,k} = \rapidity_{l,a} + ik\abs{c}$, the string center by $\rapidity_{l,a}$ and the string energy and momentum by $E_l(\rapidity_{l,a})$ and $P_l(\rapidity_{l,a})$. With this, the general wave function is given by
\begin{align}
	\Psi(t,\bs{x}) &= \qty[\prod_{l=1}^\infty\int\dd[N_l]{\rapidity_{l,\cdot}}] A(\bs{\rapidity}) \psi(\bs{x}|\bs{\rapidity}) e^{-i\sum_{l,a,k} \rapidity_{l,a,k}^2 t}.\label{equ:LL_attractive_WF}
\end{align}
Making again the choice $A(\bs{\rapidity}) = \prod_{l=1}^\infty\prod_{a=1}^{N_l}  \tilde{A}_{l,a}\qty(\tfrac{\rapidity_{l,a}-\rapidity^0_{l,a}}{\Delta \rapidity_{l,a}}) e^{-i\ell\sum_{l,a} \rapidity_{l,a} y_{l,a}^0}$, evaluating the wavefunction at large scales $x\to \ell x, t\to \ell t$ we find analogously to \eqref{equ:LL_SP_phase} the following fast oscillating phase
\begin{align}
	S_t(\bs{x},\bs{\rapidity}) &= \sum_{l,a,k} \rapidity_{l,a,k} (x_{l,a,k}-\hat{x}_{l,a}^0) +\tfrac{1}{4\ell}\sum_{(l,a,k)\neq (l',a',k')}\sgn(x_{l,a,k}-x_{l',a',k'}) \phi(\rapidity_{l,a,k}-\rapidity_{l',a',k})- \sum_{l,a,k}\rapidity_{l,a,k}^2 t.\label{equ:LL_attractive_phase_basic}
\end{align}
Now observe that due the exponential decay of the wave function of a string, the wave function will only be non-zero if all string components are at the same location $x_{l,a,k} \approx x_{l,a}$. Restricting to $x_{l,a,k} = x_{l,a}$, the fast oscillating phase becomes
\begin{align}
	S_t(\bs{x},\bs{\rapidity}) &= \sum_{l,a} P_l(\lambda_{l,a}) (x_{l,a}-\hat{x}_{l,a}^0) +\tfrac{1}{4\ell}\sum_{(l,a)\neq (l',a')}\sgn(x_{l,a}-x_{l',a'}) \phi_{ll'}(\rapidity_{l,a}-\rapidity_{l',a'})- \sum_{l,a}E_l(\rapidity_{l,a}) t,\label{equ:LL_attractive_phase_string}
\end{align}
where $\varphi_{ll'}(\rapidity) = \sum_{kk'}\phi(\rapidity +i(k-k')\abs{c})$ is the scattering phase of two strings. These are precisely the properties of strings found also in the TBA formalism. Therefore, an analogous derivation as the one leading to \eqref{equ:LL_SP_QP_def} leads to the GHD of the attractive Lieb-Liniger model (which has been obtained from the TBA here~\cite{Koch_2022}).

\section{Concavity of Bethe phase}\label{app:Bethe_phase_concave}
For performing the stationary phase approximation \eqref{equ:stationary_phase}, it is important to ensure that there exists a unique point of stationary phase. We will now show that for $t>t\ind{c}= \tfrac{N}{\ell}\sup_\rapidity\abs{\varphi'(\rapidity)}$, its hessian 
\begin{align}
	\vb{H}_{ij} &= -t \delta_{ij} + \tfrac{1}{2\ell} \delta_{ij}\sum_{k\neq i}\sgn(x_i-x_k)\varphi'(\rapidity_i-\rapidity_k) - \tfrac{1}{2L}\sgn(x_i-x_j)\varphi'(\rapidity_i-\rapidity_j).
\end{align}
is negative definite, hence a unique stationary point always exists. 

The proof of this is simple: note that we can write $\vb{H}_{ij} = -\delta_{ij} + \sum_k \vb{A}_{ik} - \vb{A}_{ij}$, where $\vb{A}_{ij} = \tfrac{1}{2\ell}\sgn(x_i-x_j)\varphi'(\rapidity_i-\rapidity_j)$. Any single matrix entry of $\abs{\vb{A}_{ij}} \leq \tfrac{1}{2\ell}\sup_\rapidity\abs{\varphi'(\rapidity)}$ is bounded. Therefore, $\vec{v}^T\vb{A}\vec{v} \leq \tfrac{N}{2L}\sup_\rapidity\abs{\varphi'(\rapidity)} \vec{v}^T\vec{v}$ and thus
\begin{align}
	\vec{v}^T\vb{H}\vec{v} \leq \qty[-t + \tfrac{N}{\ell}\sup_\rapidity\abs{\varphi'(\rapidity)}] \vec{v}^T\vec{v}. 
\end{align}
This implies that $\vb{H}$ is negative definite for $t > t\ind{c} = \tfrac{N}{\ell}\sup_\rapidity\abs{\varphi'(\rapidity)}$, hence $S_t$ is concave.

\section{Action of densities of conserved quantities on Bethe states}\label{app:densities_of_cqs}
In the main text it is proposed that the densities (see \eqref{equ:LL_deriv_cq_def})
\beq\label{qarho}
	\h q_a(x) = {\rm He}\int \dd\theta\,\theta^a\h\rho(x,\theta),
\eeq
where ${\rm He}$ means Hermitian part, are local for all $a\in\N$. 

Our proof will not be fully mathematically rigorous, due to issues with same site $\delta$ functions that need to properly regularized. A full analysis of the resulting conserved densities, and a mathematical rigorous treatment, are beyond the scope of this paper.

As in section \ref{sec:heisenberg}, we will consider the non-orthogonal wave functions \eqref{equ:unormalized_chi} on the Hilbert space $L^2(\R^N)$ and later restrict to its fully symmetric subspace.

We consider three families of operators: the usual position and momentum operators $\h x_i$ and $\h p_i$, and the rapidity operators $\h \theta_i$, defined via \eqref{deftheta}. In terms of these operators, we have the empirical density
\beq\label{rhohattheta}
	\h\rho(x,\theta) = \sum_i \delta(x-\h x_i)\delta(\theta-\h\theta_i).
\eeq

The idea of the proof is now simple: away from any $x_i=x_j$, the operators $\hat{\theta}_i$ have the same action as $\h p_i$. Hence, we have
\begin{align}
    \h q_a(x) = \frac{1}{2}\qty(\int\dd{\theta} \h\rho(x,\theta) \theta^a + \mathrm{h.c.}) =  \frac{1}{2}\qty(\sum_i \delta(x-\h x_i)\h \theta_i^a + \mathrm{h.c.}) \overset{x_i\neq x_j}{=} \frac{1}{2}\qty(\sum_i \delta(x-\h x_i)\h p_i^a + \mathrm{h.c.})
\end{align}
The rhs is clearly a local operator. However, since $\chi(\bs x|\bs\theta)$ has a jump at $x_i=x_j$, there are additional terms, which are localized at $x_i=x_j$. Since, due to the $\delta(x-x_i)$, $x_i$ is located at $x$, therefore also all $x_j$ appearing from one of these jumps are located at $x$. Hence, the whole expression is local $x$. This also shows why $\h q_a(x)$ is a local density for integer $a$ only: for non-integer $n$, already away from $x_i=x_j$, the operator $\h \theta_i$ is given by a fractional derivative (which are non-local).

In the following we carry out this idea in detail: as said above, for any given $i,j$ with $i\neq j$, the function $\chi(\bs x|\bs\theta)$, as a function of $x_i$ for fixed $x_k$, $k\neq i$, has a jump at $x_i=x_j$. Assuming that $x_j$ is isolated from other coordinates different from $i$, that is $x_k\neq x_j$ for all $k\neq i,j$, this jump is
\beq
	\Delta_{ij,\bs\theta}(\bs x) = \chi(\bs x|\bs\theta)\big|^{x_i=x_j+0^+}_{x_i=x_j-0^+} = \frac{1}{(2\pi)^{N/2}}2\cos \frc{\phi(\theta_{ij})}2 \prod_{k< l\atop \{k,l\}\neq \{i,j\}}\hspace{-0.4cm}\sgn(x_{kl})\exp\Big[\ri \bs\theta\cdot\bs x + \frc{\ri}2\sum_{k< l\atop \{k,l\}\neq \{i,j\}} \hspace{-0.4cm}\phi(\theta_{kl})\sgn(x_{kl})\Big]_{x_i=x_j}
\eeq
where $x_{kl} = x_k-x_l$, etc. The value of the jump at $x_j$ is different if, for instance, $x_k=x_j$ for some other $k\neq i,j$. Note that in this definition, $\Delta_{ij,\bs\theta}(\bs x)$ does not depend on $x_i$. Likewise, consider the symmetrised expression
\beq\begin{aligned}
	\Omega_{ij,\bs\theta}(\bs x) &= \frc12\Big(\chi(\bs x|\bs\theta)\big|_{x_i=x_j+0^+}+\chi(\bs x|\bs\theta)\big|_{x_i=x_j-0^+}\Big) \\ &= \frac{1}{(2\pi)^{N/2}}\ri \sin \frc{\phi(\theta_{ij})}2 \prod_{k< l\atop \{k,l\}\neq \{i,j\}}\hspace{-0.4cm}\sgn(x_{kl})\exp\Big[\ri \bs\theta\cdot\bs x + \frc{\ri}2\sum_{k< l\atop \{k,l\}\neq \{i,j\}} \hspace{-0.4cm}\phi(\theta_{kl})\sgn(x_{kl})\Big]_{x_i=x_j}.
	\end{aligned}
\eeq

Given $i$ and assuming that all coordinates $x_j$ with $j\neq i$ are isolated from each other, we have
\beq\label{pifirst}
	\h p_i \chi(\bs x|\bs\theta) = \theta_i\chi(\bs x|\bs\theta) -\ri \sum_{j\neq i}\delta(x_{ij}) \Delta_{ij,\bs\theta}(\bs x).
\eeq
Define the regularised delta-function
\beq
	\delta_\ep(x) = \frc1{\ep\sqrt \pi}e^{-x^2/\ep^2}.
\eeq
Then
\beq
	\lim_{\ep\to0}\delta_\ep(x_{ij}) \chi(\bs x|\bs\theta) =
	\delta(x_{ij}) \Omega_{ij,\bs\theta}(\bs x)
\eeq
(as a distribution in $x_i$). Noting that
\beq
	\frc{\Delta_{ij,\bs\theta}(\bs x)}{\Omega_{ij,\bs\theta}(\bs x)} = -2\ri \cot \frc{\phi(\theta_{ij})}2,
\eeq
we obtain from \eqref{pifirst}
\beq\label{pipsitheta}
	\h p_i \chi(\bs x|\bs\theta) = \theta_i\chi(\bs x|\bs\theta) -2\sum_{j\neq i}\cot\frc{\phi(\theta_{ij})}{2} \lim_{\ep\to0}\delta_\ep(x_{ij})\chi(\bs x|\bs\theta).
\eeq
With the understanding that the results are valid up to vanishing terms as $\ep\to0$, we will drop the explicit $\lim_{\ep\to0}$.

If some of the coordinates $x_j$ with $j\neq i$ coincide, the relation \eqref{pipsitheta} is modified by a finite change of the factor $\cot\frc{\phi(\theta_{ij})}{2}$. Under $\int \dd^N x\,f(\bs x)$, the result does not depend on this modification for any rapidly decreasing function $f(\bs x)$. As rapidly decreasing functions are dense in\footnote{One also needs them to be dense in the domain of the self-adjoint operator $\h p_i$, which they are. Note that $\chi(\bs x|\bs\theta)$ is not in that domain because it is discontinuous: we take care of this by keeping track of the delta-function terms. A fully mathematically accurate description is beyond the scope of this paper.} $L^2(\R^N)$, the result \eqref{pipsitheta} can be assumed to hold for all $\bs x$.

Using \eqref{deftheta} and $\cot\tfrac{\phi(\theta)}{2} = \tfrac{c}{\theta}$, we obtain from \eqref{pipsitheta} the fundamental operatorial equation (on the dense subspace spanned by $\chi(\bs x|\bs\theta)$'s)
\beq\label{pithetaifundamental}
	\h p_i = \h\theta_i - 2c\sum_{j\neq i}\delta_\ep(\h x_{ij})\h\theta_{ij}^{-1}.
\eeq

For every permutation $\sigma\in S_N$, we define the permutation operator $\h\sigma$ as
\beq
	\h\sigma\Psi(\bs x) = \Psi(\bs x_{\sigma}).
\eeq
Because $\chi(\bs x_\sigma|\bs\theta_\sigma) = (-1)^{|\sigma|}\chi(\bs x|\bs\theta)$, we have
\beq
	\h\sigma^{-1}\h\theta_i\h\sigma = \h\theta_{\sigma(i)}
\eeq
(note that the factor $(-1)^{|\sigma|}$ plays no role here). Clearly $\chi(\bs x|\bs\theta)$ is invariant under $\h\sigma$.

We denote $\sigma_{ij}$ the permutation $i\leftrightarrow j$. We note that we have, up to vanishing corrections as $\ep\to0$,
\beq\label{sigmadelta}
	\delta_\ep(\h x_{ij}) = \h\sigma_{ij}
	\delta_\ep(\h x_{ij})
\eeq
by symmetry of the regularised delta-function and the fact that it sets the coordinates infinitesimally close to each other in a symmetric fashion in the wave function.

We now establish the relations between powers and products of $\h p_j$'s, and powers and products of $\h\theta_i$'s, on the bosonic subspace of $L^2(\R^N)$. Let us denote by $\stackrel{\rm B}=$ operatorial equality on the bosonic subspace. Clearly $\h\sigma \stackrel{\rm B}= \1$.  As an example of calculation, from the fundamental relation \eqref{pithetaifundamental} and using \eqref{sigmadelta}, we have
\beqa
	\h p_i 
	&=&
	\h\theta_i - 2c\sum_{j\neq i}\h\sigma_{ij}\delta_\ep(\h x_{ij})\h\theta_{ij}^{-1}\n
	&=&
	\h\theta_i - 2c\sum_{j\neq i}\delta_\ep(\h x_{ji})\h\theta_{ji}^{-1}\h\sigma_{ij}\n
	&\stackrel{\rm B}=&
	\h\theta_i + 2c\sum_{j\neq i}\delta_\ep(\h x_{ij})\h\theta_{ij}^{-1}.
\eeqa
Therefore
\beq
	\h p_i \stackrel{\rm B}= \h \theta_i.
\eeq
This means that the total momentum density
\begin{align}
    \h q_1(x) &= \sum_i \delta(x-\h x_i)\h p_i \stackrel{\rm B}= \sum_i \delta(x-\h x_i)\h \theta_i
\end{align}
is indeed a local density.

We proceed by induction. We show that for all subsets $I$ of $\{1,2,\ldots,N\}$ with multiplicities, we have
\beq\label{thetap}
	\h\theta_I \stackrel{\rm B}= \h p_I + \h \ell_I
\eeq
where $\theta_I = \prod_{i\in I}\h\theta_i$, $p_I = \prod_{i\in I}\h p_i$. Here $\h \ell_I$ is ``local at $\{x_i:i\in I\}$". More precisely, this means that every term in $\h\ell_I$ is formed of a product of (regularised) delta-functions involving coordinates $x_j,\,j\in J$ for some set $J$, that restrict all such $x_j$'s to lie at (i.e.~within a small neighbourhood of) at least one of the coordinates $\h x_i$, $i\in I$; right-multiplied by the momenta $\h p_j$, $j\in I\cup J$ and their powers. This indeed guarantees that $\sum_j \delta(x-\h x_j)\h\theta_j^a$ is local at $x$ for all $a=0,1,2,3,\ldots$.

Let us assume that \eqref{thetap} holds for all $I$ of order $|I|\leq n$ (counting multiplicities). Clearly \eqref{thetap} indeed holds for $n=0,1$. Note that for every $j$, we have that $\h p_j \h \ell_I$ is local for $\{j\}\cup I$. Choose $I$ of order $n$, and write $\h\theta_I = \prod_j \h\theta_j^{m_j}$. Then
\beqa
	\h\theta_{\{i\}\cup I} &=& \Big(\h p_i + 2c\sum_{j\neq i}\delta_\ep(\h x_{ij})\h\theta_{ij}^{-1}\Big) \h\theta_I\n
	&=& \h p_i\h p_I + 2c\sum_{j\neq i}\delta_\ep(\h x_{ij})\h\theta_{ij}^{-1}\h\theta_i^{m_i}\h\theta_j^{m_j}\prod_{k\neq i,j}\h\theta_k^{m_k} + \h p_i\h \ell_I \n
	&=&\h p_{\{i\}\cup I}+ 2c\sum_{j\neq i}\frc{1+\h\sigma_{ij}}2\delta_\ep(\h x_{ij})\h\theta_{ij}^{-1}\h\theta_i^{m_i}\h\theta_j^{m_j}\prod_{k\neq i,j}\h\theta_k^{m_k}  + \h p_i\h \ell_I \n
	&\stackrel{\rm B}=&  \h p_{\{i\}\cup I}  + c\sum_{j\neq i}\delta_\ep(\h x_{ij})p_{m_i,m_j}(\h\theta_i,\h\theta_j)\prod_{k\neq i,j}\h\theta_k^{m_k} + \h p_i\h \ell_I
    \label{thetahatp}
\eeqa
where we define the polynomial $p_{m,m'}(\theta,\theta') = (\theta^m{\theta'}^{\,m'} - \theta^{m'}{\theta'}^{\,m})/(\theta-\theta')$, of order $m+m'-1$. The polynomial $p_{m_i,m_j}(\h\theta_i,\h\theta_j)\prod_{k\neq i,j}\h\theta_k^{m_k}$ in \eqref{thetahatp} is of order $|I|-1$, therefore we can use recursion (by steps of 2 in $|I|$), and we find $\h\theta_{\{i\}\cup I} = \h p_{\{i\}\cup I} + \ell_{\{i\}\cup I}$ for appropriate $\ell_{\{i\}\cup I}$. Thus \eqref{thetap} holds for all $I$.

A proper treatment of the regularisation is necessary. In evaluating $\h\theta^a$ for $a\leq 3$ no ambiguities arise. We find,
\begin{align}\label{thetai1}
    \hat{\theta}_i &\stackrel{\rm B}= \hat{p}_i\\ \label{thetai2}
    \hat{\theta}_i^2 &\stackrel{\rm B}= \hat{p}_i^2 + c \sum_{j\neq i}\delta_\epsilon(\hat{x}_{ij})\\
    \hat{\theta}_i^3 &\stackrel{\rm B}= \hat{p}_i^3 + 3c \sum_{j\neq i}\delta_\epsilon(\hat{x}_{ij}) \hat{p}_i -\ri c \sum_{j\neq i}\delta'_\epsilon(\hat{x}_{ij})\label{equ:densities_rel_3}\\
    \h\theta_i\h\theta_j &\stackrel{\rm B}= \h p_i\h p_j - c\delta_\ep(\h x_{ij})\quad (i\neq j).
\end{align}
Note that \eqref{thetai2} implies that the energy density
\begin{align}
    \h q_2(x) = \sum_i \delta(x-\h x_i) \qty(\hat{p}_i^2 + c \sum_{j\neq i}\delta_\epsilon(\hat{x}_{ij}))
\end{align}
is a local density.
Similarly, we can use \eqref{equ:densities_rel_3} to explicitly find
\beqa
    \h p_i^3 &\stackrel{\rm B}=&
    \h p_i \Big(\h\theta_i^2 - c\sum_{j\neq i}\delta_\ep(\h x_{ij}\Big)\qquad\mbox{(Eq.~\eqref{thetai2})}\n&=&
    \h p_i\h\theta_i^2 - c\sum_{j\neq i}\delta_\ep(\h x_{ij})\h p_i + \ri c \sum_{j\neq i}\delta_\ep'(\h x_{ij})\n
    &=& \Big(\h\theta_i - 2c\sum_{j\neq i}\delta_\ep(\h x_{ij})\h\theta_{ij}^{-1}\Big)\h\theta_i^2 - c\sum_{j\neq i}\delta_\ep(\h x_{ij})\h p_i + \ri c \sum_{j\neq i}\delta_\ep'(\h x_{ij})
    \qquad\mbox{(Eq.~\eqref{pithetaifundamental})}\n
    &\stackrel{\rm B}=&
    \h\theta_i^3 - c\sum_{j\neq i}\delta_{\ep}(\h x_{ij})
    (\h\theta_i+\h\theta_j)
    - c\sum_{j\neq i}\delta_\ep(\h x_{ij})\h p_i + \ri c \sum_{j\neq i}\delta_\ep'(\h x_{ij})\qquad\mbox{(symmetry)}\n
    &\stackrel{\rm B}=&
    \h\theta_i^3 - 2c\sum_{j\neq i}\delta_{\ep}(\h x_{ij})
    \h\theta_i
    - c\sum_{j\neq i}\delta_\ep(\h x_{ij})\h p_i + \ri c \sum_{j\neq i}\delta_\ep'(\h x_{ij})\qquad\mbox{(symmetry)}\n
    &\stackrel{\rm B}=&
    \h\theta_i^3 - 3c\sum_{j\neq i}\delta_{\ep}(\h x_{ij})
    \h p_i
    + \ri c \sum_{j\neq i}\delta_\ep'(\h x_{ij})\qquad\mbox{(Eq.~\eqref{thetai1})}
\eeqa
where ``symmetry" means the use of \eqref{sigmadelta} along with $\h\sigma \stackrel{\rm B}= \1$. From this find the density $\h q_3(x)$ of the third conserved quantity of the Lieb-Liniger model
\begin{align}
    \h q_3(x) = \He\sum_i \delta(x-\h x_i)\Big(
 	\h p_i^3 + 3c \sum_{j\neq i}
 	\delta(\h x_i-\h x_j)\h p_i\Big).
\end{align}
This coincides with the result obtained in~\cite{davieshigher,davieskorepin}.

However, for $a\geq 4$, one seemingly obtains squares and higher powers of delta functions. The above calculation provides an unambiguous way of treating such terms: in \eqref{thetahatp}, the limits $\ep\to0$ involved in evaluating the polynomial $p_{m_i,m_j}(\h\theta_i,\h\theta_j)\prod_{k\neq j}\h\theta_k^{m_k}$ must be taken before the limit on the factor $\delta_{\ep}(\h x_{ij})$. By construction, any polynomial in $\h\theta_i$'s is regular, hence the result is finite. The results are expressions involving multiple regularisation parameters, with limits that must be taken in the appropriate order. The fact that a regular operator is obtained by an expression formally involving powers of delta functions is carefully explained in \cite{davieskorepin}. As an example, we obtain, re-introducing the limit symbols,
\beqa
    \h \theta_i^4 &\stackrel{\rm B}=& \h p_i^4
    + \lim_{\ep\to0}c\sum_{j\neq i}\delta_\ep(\h x_{ij})
    \lim_{\ep'\to0} \Big(\h p_i^2 + \h p_i\h p_j + \h p_j^2
    + c\sum_{k\neq i,j}\big(\delta_{\ep'}(\h x_{ik}) + \delta_{\ep'}(\h x_{kj})\big) +c\delta_{\ep'}(\h x_{ij})\Big)\n
    && +\,\lim_{\ep\to0}\sum_{j\neq i}(3c\delta_\ep(\h x_{ij})\h p_i^2
    + 4c\ri \delta'_\ep(\h x_{ij})\h p_i
    + c \delta''_\ep(\h x_{ij})).
\eeqa
Note how on the first line, there is indeed a term proportional to $\delta_{\ep}(\h x_{ij})\delta_{\ep'}(\h x_{ij})$.

\section{The Bethe scattering map $x\to y$}\label{app:scattering_map}

We first analyse the Bethe scattering map $\bs y^{\bs \theta}:\bs x\to \bs y$, Eq.~\eqref{ythetamap}, mapping real coordinates $\bs x$ to free coordinates $\bs y$, as well as its inverse map $\bs x^{\bs\theta}:\bs y\to\bs x$. A simple initial remark is that the map preserves the ``center of mass'':
\begin{equation}
    \sum_i y_i = \sum_i x_i.
\end{equation}

In general, {\em the function $\bs y^{\bs\theta}$ is not surjective}. In fact, because of the sign function involved, it is well defined only for real coordinates $\bs x$ in the ``regular domain" $\R^N_{\neq}:=\{\bs x\in\R^N:x_i\neq x_j\;\forall\;i\neq j\}$, which avoids all diagonals. This domain is the disconnected union of $N!$ continua, $\R^N = \cup_{\sigma\in S_N} \R^N_{\sigma}$, which are the open ``quadrants" associated to different orderings of the coordinates, $\R^N_{\sigma} = \{\bs x \in\R^N: x_{\sigma_{i+1}}>x_{\sigma_i}\;\forall\;i=1,2,\ldots N-1\}$. The image of $\bs y^{\bs\theta}$ on $\R^N_{\neq}$ is what we will refer to as the ``permitted region" of values of the free coordinates $\bs y$. It takes the form of the union of well separated continua $\R^N_{\not\equiv}:=\cup_{\sigma\in S_N} R^N_{\sigma}$, which are the quadrants $\R^N_\sigma$ shifted by constants:
\begin{equation}
    R^N_\sigma = \R^N_\sigma + 
    \bs r^\sigma,\quad r^\sigma_{\sigma_i} = \frc12 \sum_{j\neq i} \varphi(\theta_{\sigma_i}-\theta_{\sigma_j}) \,\sgn(i-j).
\end{equation}
That is,
\begin{equation}
    y^{\bs\theta}(\R^N_{\neq}) = \R^N_{\not\equiv}.
\end{equation}

Surjectivity can be recovered in various ways. One may regularise the sign function, say
\begin{equation}\label{regulari}
    \sgn(x) \to \tanh(x/\ep)\quad
    \mbox{(regularised Bethe scattering map)}
\end{equation}
for some $\ep>0$ small enough, whereby $\bs x\mapsto \bs y$ becomes a diffeomorphism. The clearest way, however, is to take the limit $\ep\to0$ from a regularisation -- the ``renormalisation" of the map. This is largely independent of the regularisation chosen, and the result is that {\em the map $\bs y^{\bs\theta}$ becomes surjective, at the price of it not being a function anymore}: in Eq.~(9), the image of 0 under the sign function is simply defined as the interval
\begin{equation}\label{signintverval}
    \sgn(0) = [-1,1]\quad
    \mbox{(renormalised Bethe scattering map, $\bs x\mapsto \bs y$)}.
\end{equation}
From now on, we will take this definition. Then $\bs y^{\bs\theta}$ is well defined on the ``singular set" $\R^N_= := \R^N\setminus \R^N_{\neq}$ as well. We will refer to the complement of the permitted region, $\R^N_{\equiv}:=\R^N\setminus \R^N_{\not\equiv}$, as the ``forbidden region", the image of $\bs y^{\bs \theta}$ on the singular set,
\begin{equation}
    y^{\bs\theta}(\R^N_{=}) = \R^N_{\equiv}.
\end{equation}
In particular, the image of any point $\bs x\in \R^N_=$ is its ``connecting set" in $\R^N_\equiv$, which connects (identifies) points on the boundary of $\R^N_\equiv$.

\begin{figure}
\bc\includegraphics[height= 6cm]{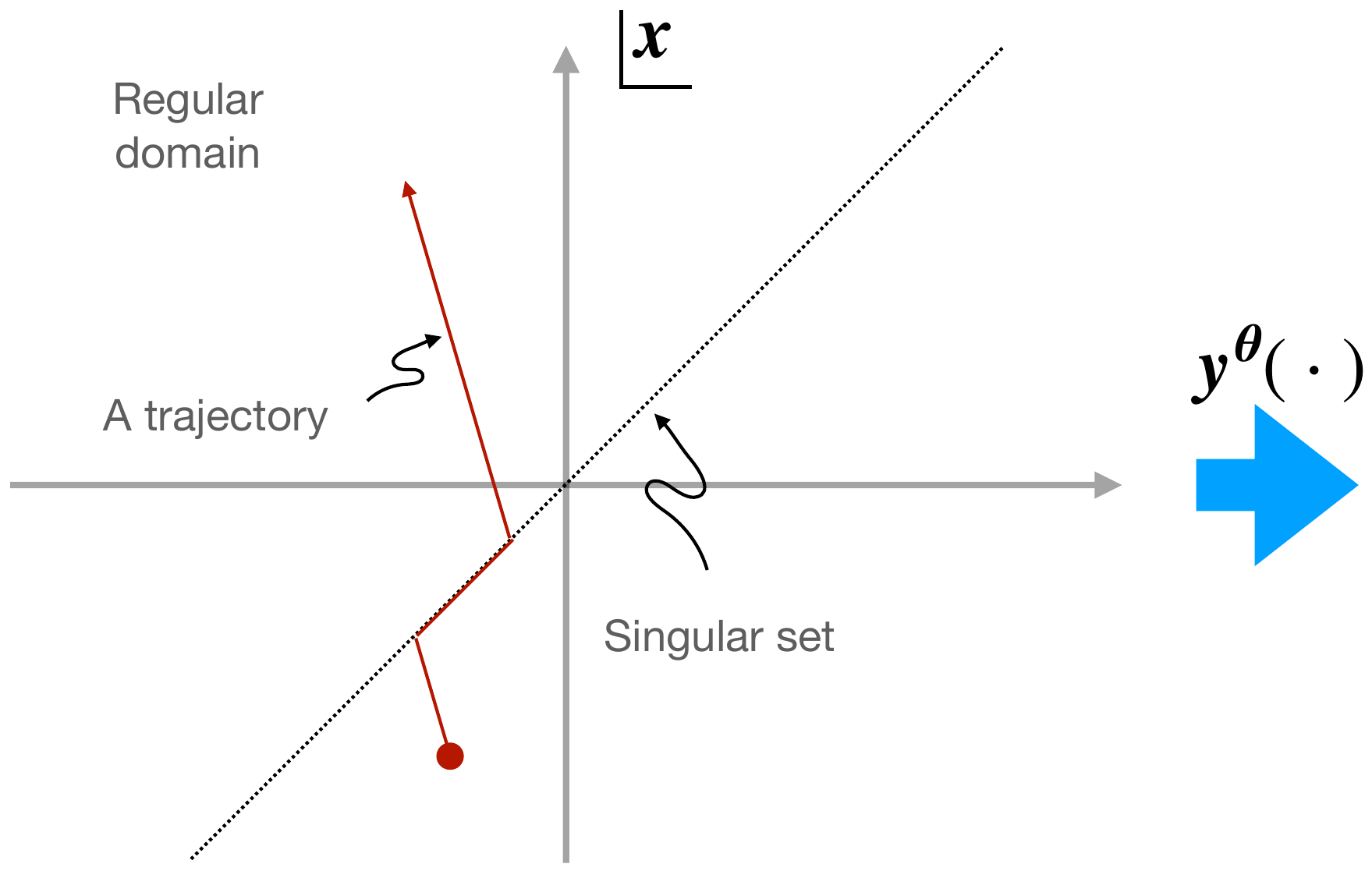}\includegraphics[height= 6cm]{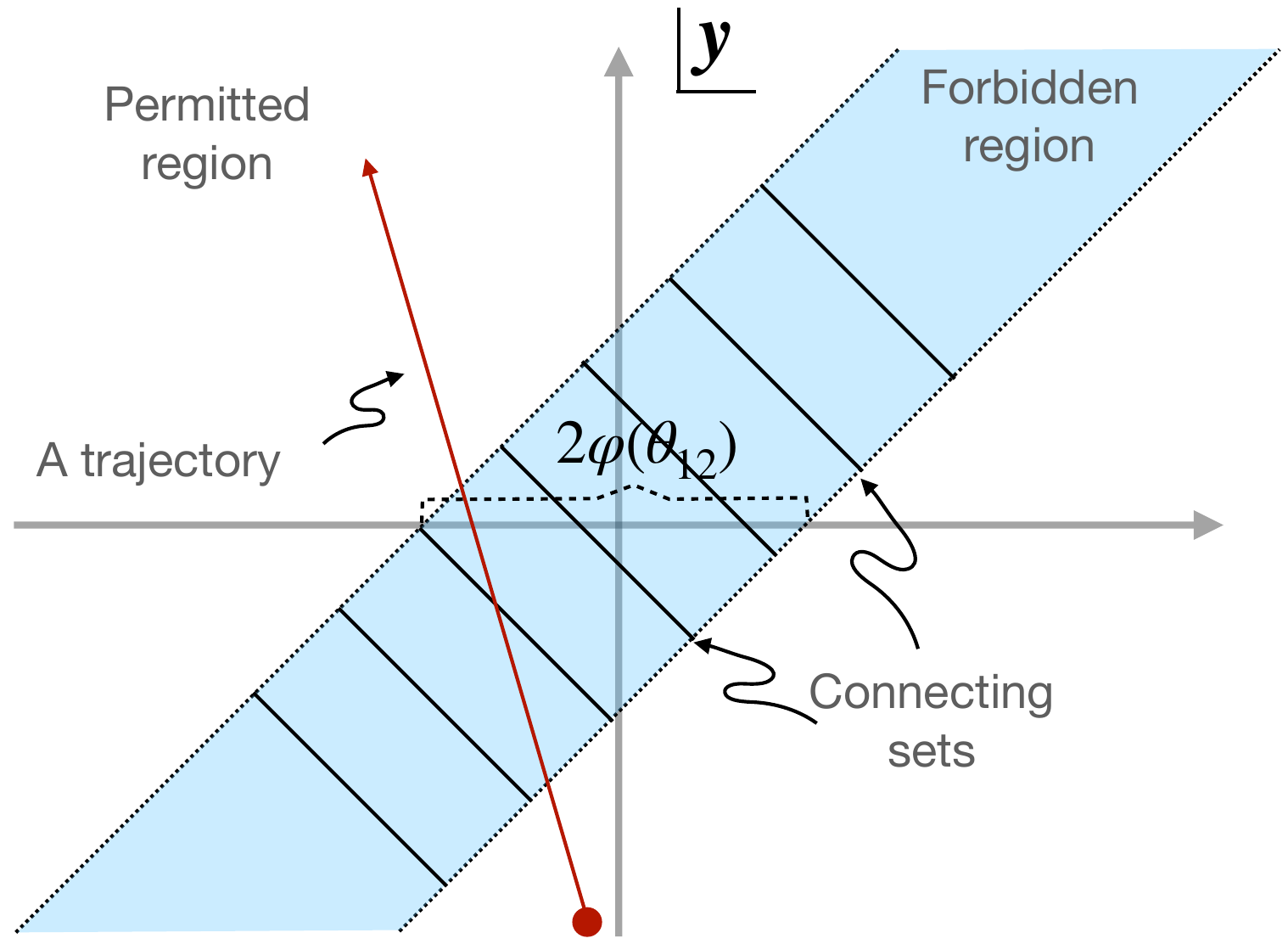}\ec
\caption{The mapping $\bs x\mapsto \bs  y = \bs y^{\bs\theta}(\bs x)$ from Eq.~\eqref{ythetamap}, in its renormalised version with $\sgn(0) = [-1,1]$, for the case of two particles. The singular set maps onto the forbidden region of the $\bs y$ plane, each point mapping to the corresponding connecting set. A trajectory is shown: in real coordinates ($\bs x$), particles stick to each other, for the time that their straight trajectory in free coordinates ($\bs y$) spends within the forbidden region.}
\label{figyplane}
\end{figure}

As an illustration, in the case $N=2$ we have
\begin{equation}
    \bs y^{\bs \theta} \big(\{(x_1,x_2):x_1\neq x_2\} = \R^2_{\neq}\big) = \R^2_{\not\equiv} = \{(y_1,y_2):|y_1-y_2|>\varphi(\theta_{12})\}
\end{equation}
as shown in Fig.~\ref{figyplane}.

On every quadrant $\R^N_\sigma$, the map $\bs y^{\bs\theta}$ is an affine function: it is simply a shift by a quadrant-dependent constant,
\begin{equation}\label{displacement}
    \bs y = \bs x + \bs r^\sigma,\quad
    \bs x\in\R^N_\sigma.
\end{equation}
The inverse map $\bs x^{\bs\theta}$ is likewise, on every connected component $R_\sigma^N$ of the permitted region $\R_{\not\equiv}^N$, an affine function.

On the singular set, as mentioned $\bs y^{\bs\theta}$ it is not a function, as it maps every point to a nontrivial subset (which may be an interval or a higher-dimensional region) of the forbidden region. However, it is surjective, and therefore $\bs x^{\bs\theta}$ is well defined on $\R^N$, including on the forbidden region $\R_{\equiv}^N$. The result can be expressed as the minimisation problem Eq.~(14). This  defines $\bs x^{\bs\theta}$ unambiguously, including on $\R^N_{\equiv}$, which maps onto $\R^N_=$ (surjectively, but not injectively). Then, $\bs x^{\bs\theta}$ is a piecewise affine function throughout $\R^N = \R^N_{\not\equiv}\cup\R^N_{\equiv}$, including throughout the forbidden region $\R^N_{\equiv}$. More precisely, it behaves as follows.

For $\bs y$ lying either within or outside the forbidden region, if $x_i$ does not coincide with any other coordinate, then it still satisfies $\p_{y_j}x_i = \delta_{ij}\;\forall\;j$. However, suppose there is a cluster $I$, i.e.~a subset $I\subset\N$ (with $|I|\geq 2$ elements) of coordinates that coincide, $x_i = x_j\;\forall\;i,j\in I$, but that are distinct from all other coordinates. Then one makes in Eq.~(9) the replacement
\begin{equation}
    \sgn(x_i-x_j)\to f_{ij}(\bs y_{I})
\end{equation}
for all $i,j\in I,\;i\neq j$, where $\bs y_{I} = (y_i)_{i\in I}$. Here, $f_{ij}(\bs y_I) = -f_{ji}(\bs y_I)$ are appropriate affine real-valued functions, whose forms only depend on $I$ and on the set of coordinates $J:=\{j:x_j>x_i\;\forall \;i\in I\}$ lying to the right of $I$ (or equivalently, the set of coordinates to its left). That is, one writes
\begin{align}\label{xrenormalised}
    x_i = y_i - \sum_{j\in I,\,j\neq i}\varphi(\theta_{ij})
    f_{ij}(\bs y_I) + \sum_{j\in J} \varphi(\theta_{ij})
    -
    \sum_{j\not\in I\cup J} \varphi(\theta_{ij}),\quad i\in I,\quad \mbox{$I$ a cluster, $\bs y\in \R^N_{\equiv}$.}&\\
    \mbox{(renormalised Bethe scattering map, $\bs y\mapsto \bs x$).}&
\end{align}
It is possible to work out the affine functions $f_{ij}(\bs y_I) = \sum_{k\in I} a_{ijk} y_k + c_{ij}$ explicitly. For a cluster of
\begin{equation}
    n=|I|
\end{equation}
elements, there are $(n+1)n(n-1)$ coefficients ($n^2(n-1)$ coefficients $a_{ijk}$ and $n(n-1)$ coefficients $c_{ij}$). The conditions to solve are $x_i = x_j$ for all $i,j\in I,\;i\neq j$ and all $\bs y_I$ such that $\bs y\in \R^N_{\equiv}$ with fixed $I,J$. There are $n(n-1)$ equations, and each equation is affine in $\bs y_I$, so we only have to put to zero the coefficient of each $y_i,\;i\in I$ and the constant: $n+1$ constraints. Thus there are $(n+1)n(n-1)$ constraints, hence the solution is unique. One can immediately write the solution (up to the constants $c_{ij}$ which we do not evaluate here):
\begin{equation}
    f_{ij}(\bs y_I) = \frac{y_i-y_j}{n\varphi(\theta_{ij})} + c_{ij}
\end{equation}
as then
\begin{equation}
    y_i - y_j - \sum_{k\in I,\,k\neq i}\varphi(\theta_{ik})
    f_{ik}(\bs y_I)
    +
    \sum_{k\in I,\,k\neq j}\varphi(\theta_{jk})
    f_{jk}(\bs y_I)
    = const.
\end{equation}
is a constant for every $i,j\in I$ (and requiring $x_i-x_j=0$ then fixes the $c_{ij}$'s). The explicit form of the coordinates in the cluster, up to a constant that depends on $i, I, J$, is then
\begin{equation}
    x_i = \frc{\sum_{j\in I} y_j}{n} + const.,
    \quad \mbox{$I$ a cluster, $\bs y\in \R^N_{\equiv}$.}
\end{equation}
That is, the real coordinates in a cluster simply are the arithmetic averages of the free coordinates (up to additive constants). The slopes are therefore
\begin{equation}
    \p_j x_i = \frc1n,\quad i,j\in I,
    \quad\mbox{$I$ a cluster, $\bs y\in \R^N_{\equiv}$.}
\end{equation}

Finally, considering the dynamics, Eq.~\eqref{equ:LL_SP_QP_def_micro}, it is clear that whenever $x_i$ is separated from every other coordinates it evolves with velocity $\theta_i$, and that all particles in a cluster $I$ evolve with the cluster's average velocity,
\begin{equation}
    \dot x_i = \frc{\sum_{j\in I}\theta_j}{n},\quad i\in I,\quad
    \mbox{$I$ a cluster, $\bs y\in \R^N_{\equiv}$.}
\end{equation}
Note how the velocity of a cluster does not depend on the interaction $\varphi(\theta)$; however, the time each coordinate spends within a cluster depends on it, as the set of $\varphi(\theta_{ij})$'s determines the various widths of the forbidden region  $\R^N_{\equiv}$ (along various coordinates) via the displacements \eqref{displacement}.

It is instructive to look at the case $N=2$. Within the permitted region, the inverse map is
\begin{equation}
    x_1 = y_1 - s\frc{\varphi(\theta_{12})}2,\quad
    x_2 = y_2 + s\frc{\varphi(\theta_{12})}2,\quad
    (y_1,y_2)\in \R^2_{\not\equiv},\  s=\sgn(y_1-y_2).
\end{equation}
Within the forbidden region $\R^2_{\equiv}$, according to the above discussion the inverse map is obtained from the replacement $\sgn(x_1-x_2)\to a y_1 + b y_2 + c$ and takes the general form
\begin{equation}
    x_1=x_1^{\bs \theta}(y_1,y_2)
    = y_1 - \frc{\varphi(\theta_{12})}2(a y_1 + b y_2 + c),\quad
    x_2=x_2^{\bs \theta}(y_1,y_2)
    = y_2 + \frc{\varphi(\theta_{12})}2(a y_1 + b y_2 + c).
\end{equation}
The constants $a,b,c$ are fixed by the requirement that $(x_1,x_2)$ lie on the singular domain:
\begin{equation}
    x_1 = x_2 \;\Rightarrow\;
    y_1 - y_2 = \varphi(\theta_{12})(ay_1+by_2+c).
\end{equation}
The solution is simply $a = -b = 1/\varphi(\theta_{12})$, $c=0$, giving
\begin{equation}
    x_1 =x_2 = \frc{y_1+y_2}2,\quad
    (y_1,y_2) \in \R^N_{\equiv}
\end{equation}
which of course could have been deduced directly from preservation of the center of mass. Here, the lifetime of the cluster under the dynamics Eq.~\eqref{equ:LL_SP_QP_def_micro} is $t = t_+-t_-$ obtained from (assuming $\theta_1>\theta_2$)
\begin{equation}
    a(y_1 + \theta_1 t_\pm) + b(y_2 + \theta_2 t_\pm) + c = \pm 1\ \Rightarrow\ 
    t_{\pm} = \frc{\pm\varphi(\theta_{12})- y_{12}}{\theta_{12}}
\end{equation}
and thus
\begin{equation}
    t = \frc{2\varphi(\theta_{12})}{\theta_{12}}.
\end{equation}
We note that the displacement of particle 1 during this time is $\varphi(\theta_{12})\frc{\theta_1+\theta_2}{\theta_1-\theta_2}$, as compared to that of its non-interacting trajectory $\varphi(\theta_{12})\frc{2\theta_1}{\theta_1-\theta_2}$, the difference of the latter with the former being the scattering shift $\varphi(\theta_{12})$, as it should.

\printbibliography  

\end{document}